\documentclass[12pt]{article}
\usepackage{axodraw}

\makeatletter
\newcommand{\eqref}[1]{(\ref{#1})}
\newcommand{\Secref}[1]{\expandafter\MakeUppercase\secrefname~\ref{#1}}
\newcommand{\Appref}[1]{\expandafter\MakeUppercase\appendixname~\ref{#1}}
\newcommand{\Figref}[1]{\expandafter\MakeUppercase\figurename~\ref{#1}}
\newcommand{\Tbref}[1]{\expandafter\MakeUppercase\tablename~\ref{#1}}
   \let\secref\Secref  
  
\newcommand{\secrefname}{Section}
\DeclareSymbolFont{AMSb}{U}{msb}{m}{n}
\DeclareSymbolFontAlphabet{\mathbb}{AMSb}
\DeclareMathAlphabet\mathfrak{U}{euf}{m}{n}
\SetMathAlphabet\mathfrak{bold}{U}{euf}{b}{n}
\DeclareSymbolFont{boldletters}  {OML}{cmm} {b}{it}
\DeclareSymbolFont{boldoperators}{OT1}{cmr}{bx}{n}
\DeclareSymbolFont{boldsymbols}  {OMS}{cmsy}{b}{n}
\newcommand{\Vev}[1]{\Bigl\langle #1 \Bigr\rangle}
\newcommand{\inv}{^{\raise.15ex\hbox{${\scriptscriptstyle -}$}\kern-.05em 1}} 
\newcommand{\ext}[1][]{\mathop{\raisebox{.2ex}{$\textstyle\bigwedge$}^{#1}}}
\newcommand{\del}{\partial}
\newcommand{\e}{\mathrm{e}}                               
\newcommand{\dint}[2][]{\mathop{\mathalpha{\int#1}#2}}    
\newcommand{\doint}[2][]{\mathop{\mathalpha{\oint#1}#2}}  
\newcommand{\A}{\mathbb{A}}
\newcommand{\set}[1]{\mathbb{#1}}                         
\newcommand{\R}{\set{R}}

\newcommand{\group}[1]{\mathop{\kern\z@\mathrm{#1}}\nolimits}     

\newcommand{\lie}[1]{\mathop{\kern\z@\mathfrak{#1}}\nolimits}
\newcommand{\g}{\lie{g}}
\newcommand{\opname}[1]{\mathop{\kern\z@\mathrm{#1}}\nolimits}    
\newcommand{\im}{\opname{im}}                             
\newcommand{\Hom}{\opname{Hom}}
\newcommand{\eps}{\epsilon}
\newcommand{\CA}{\mathcal{A}}
\newcommand{\CB}{\mathcal{B}}
\newcommand{\CC}{\mathcal{C}}
\newcommand{\CD}{\mathcal{D}}
\newcommand{\CL}{\mathcal{L}}
\newcommand{\CM}{\mathcal{M}}
\newcommand{\CO}{\mathcal{O}}
\newcommand{\CV}{\mathcal{V}}
\DeclareMathSymbol{\Ba}{\mathalpha}{boldletters}{`a}
\DeclareMathSymbol{\Bb}{\mathalpha}{boldletters}{`b}
\DeclareMathSymbol{\Bc}{\mathalpha}{boldletters}{`c}
\DeclareMathSymbol{\Bd}{\mathalpha}{boldletters}{`d}
\DeclareMathSymbol{\Bf}{\mathalpha}{boldletters}{`f}
\DeclareMathSymbol{\Bg}{\mathalpha}{boldletters}{`g}
\DeclareMathSymbol{\Bx}{\mathalpha}{boldletters}{`x}
\DeclareMathSymbol{\By}{\mathalpha}{boldletters}{`y}
\DeclareMathSymbol{\Bz}{\mathalpha}{boldletters}{`z}
\DeclareMathSymbol{\BF}{\mathalpha}{boldletters}{`F}
\DeclareMathSymbol{\BG}{\mathalpha}{boldletters}{`G}
\DeclareMathSymbol{\BQ}{\mathalpha}{boldletters}{`Q}
\DeclareMathSymbol{\BX}{\mathalpha}{boldletters}{`X}
\DeclareMathSymbol{\Bal}{\mathord}{boldletters}{"0B}
\DeclareMathSymbol{\Bbe}{\mathord}{boldletters}{"0C}
\DeclareMathSymbol{\Bga}{\mathord}{boldletters}{"0D}
\DeclareMathSymbol{\Bde}{\mathord}{boldletters}{"0E}
\DeclareMathSymbol{\Bxi}{\mathord}{boldletters}{"18}
\DeclareMathSymbol{\Bphi}{\mathord}{boldletters}{"1E}
\DeclareMathSymbol{\Bch}{\mathord}{boldletters}{"1F}
\DeclareMathSymbol{\Bps}{\mathord}{boldletters}{"20}
\DeclareMathSymbol{\Bom}{\mathord}{boldletters}{"21}
\DeclareMathSymbol{\BPs}{\mathalpha}{boldoperators}{"09}
\DeclareMathSymbol{\BV}{\mathord}{boldsymbols}{"34}
\DeclareMathDelimiter{\Blbracket}{\mathopen} {boldoperators}{"28}{largesymbols}{"00}
\DeclareMathDelimiter{\Brbracket}{\mathclose}{boldoperators}{"29}{largesymbols}{"01}
\newcommand{\Bracket}[1]{\Blbracket #1 \Brbracket}
\renewcommand{\section}{\@startsection{section}{1}{\z@}%
                                    {-7ex \@plus -1ex \@minus -.2ex}%
                                    {2.5ex \@plus.2ex}%
                                    {\normalfont\large\scshape\centering}}
\renewcommand{\subsection}{\@startsection{subsection}{2}{\z@}%
                                       {-5ex \@plus -1ex \@minus -.2ex}%
                                       {1.5ex \@plus.2ex}%
                                       {\normalfont\normalsize\scshape}}
\newcommand{\sectionname}{}

\renewcommand\@seccntformat[1]{\ignorespaces\csname #1name\endcsname\space
                               \csname the#1\endcsname.\quad}   
\renewcommand{\appendix}{\par
  \setcounter{section}{0}%
  \setcounter{subsection}{0}%
  \renewcommand{\thesection}{\@Alph\c@section}%
  \renewcommand{\sectionname}{\appendixname}}
\newdimen\captionmargin 
\newdimen\captionindent 
\newdimen\captionwidth 
\newcommand{\captionfont}{\slshape}
\newcommand\@captionlabel[1]{\textsc{#1:}\space}
\long\def\@makecaption#1#2{%
  \vskip\abovecaptionskip  
  \captionwidth\hsize 
  \advance\captionwidth -2\captionmargin
  \sbox\@tempboxa{\@captionlabel{#1}\captionfont #2}%
  \ifdim \wd\@tempboxa >\captionwidth
    \ifdim\captionindent>\z@ 
      \advance\captionwidth -\captionindent
      \hskip\captionindent
    \fi
    \hskip\captionmargin
    \parbox[t]{\captionwidth}{\leavevmode\hskip-\captionindent
      \@captionlabel{#1}\captionfont #2}%
  \else
    \global \@minipagefalse
    \hb@xt@\hsize{\hfil\box\@tempboxa\hfil}%
  \fi
  \vskip\belowcaptionskip}
\def\eqnarray{%
   \stepcounter{equation}%
   \def\@currentlabel{\p@equation\theequation}%
   \global\@eqnswtrue
   \m@th
   \global\@eqcnt\z@
   \tabskip\@centering
   \let\\\@eqncr
   $$\everycr{}\halign to\displaywidth\bgroup
       \hskip\@centering$\displaystyle\tabskip\z@skip{##}$\@eqnsel
      &\global\@eqcnt\@ne$\;\hfil{##}$\hfil
      &\global\@eqcnt\tw@$\;\displaystyle{##}$\hfil\tabskip\@centering
      &\global\@eqcnt\thr@@ \hb@xt@\z@\bgroup\hss##\egroup
         \tabskip\z@skip
      \cr
}
\ifcase \@ptsize
\or 
\or 
\fi
  \divide\@tempdima\baselineskip
  \@tempcnta=\@tempdima
\makeatother 
\begin{document}

%
%
\thispagestyle{empty}

\begin{flushright}\scshape
YITP-SB-02-41, RUNHETC-2002-32\\
hep-th/0209148\\
September 2002
\end{flushright}
\vskip1cm

\begin{center}

{\LARGE\scshape
Topological Open Membranes
\par}
\vskip15mm

\textsc{Christiaan Hofman$^{1,\dagger}$} and 
\textsc{Jae-Suk Park$^{2,3,\ddagger}$}
\par\bigskip
{\itshape
  ${}^1$New High Energy Theory Center, Rutgers University,\\
        Piscataway, NJ 08854, USA,\\
  \par\medskip
  ${}^2$Department of Physics, KAIST,\\
        Taejon, 305-701, Korea,
  \par\medskip
  ${}^3$C.N. Yang Inst. for Theoretical Physics \textnormal{and} Dept. of Mathematics,\\
  SUNY at Stony Brook, NY 11794, USA,}
\par\bigskip
\texttt{${}^\dagger$hofman@physics.rutgers.edu, ${}^\ddagger$jaesuk@muon.kaist.ac.kr}

\end{center}

\section*{Abstract}

We study topological open membranes of BF type in a manifest BV formalism. 
Our main interest is the effect of the bulk deformations on the 
algebra of boundary operators. This forms a homotopy Lie algebra, 
which can be understood in terms of a closed string field theory. 
The simplest models are associated to quasi-Lie bialgebras and 
are of Chern-Simons type. More generally, the induced structure 
is a Courant algebroid, or ``quasi-Lie bialgebroid'', 
with boundary conditions related to Dirac bundles. 
A canonical example is the topological open membrane coupling to a 
closed 3-form, modeling the deformation of strings by a C-field. 
The Courant algebroid for this model describes a modification 
of deformation quantization. We propose our models as a tool to find 
a formal solution to the quantization problem of Courant algebroids.

\newpage
\setcounter{page}{1}
%
%

\section{Introduction}
\label{sec:intro}

Topological field theories have emerged as an important tool for 
performing exact calculations in physics. They are also very well fit to 
apply field theoretical methods to mathematical problems. 
The topological Poisson sigma model introduced in \cite{schastro} 
has been used in \cite{cafe} to give the solution of Kontsevich 
\cite{kon1,kon2} to the problem of deformation quantization 
in terms of correlation functions for a topological string theory. 
This model captures the essence of the emergence of noncommutative 
geometry in open string theory in the presence of a 
$B$-field background \cite{codo,scho,seiwit}.
This topological model, which is of BF-type, is most 
succinctly formulated in a BV language. In this 
formulation it can be viewed as a sigma-model with 
a particular symplectic superspace as target manifold. 
Many 2-dimensional topological field theories---such as the 
A- and the B-model \cite{alkoschza,js}---can be formulated 
in such a way. These topological models put on the disc 
describe the deformation theory of the algebra of boundary 
operators \cite{homaOS,hof}. In the case of \cite{cafe} the boundary algebra 
was simply the algebra of functions $C^\infty(M)$ on some manifold, 
which was quantized to a noncommutative algebra by a 
Poisson bivector coupling to the bulk. 

This idea has a straightforward generalization to higher dimensions. 
Many higher dimensional topological field theories can be formulated 
as gauge fixed versions of similar BF type BV sigma models, 
e.g.\ Chern-Simons theory, 
Rozanski-Witten theory \cite{rozwit}, Donaldson-Witten theory 
\cite{don,witdon}, and the membrane coupling to a 3-form \cite{js}. 
This gives rise to 
open $p$-branes as introduced in \cite{js}, which have 
various applications to both physics and mathematics. 
In this paper we will specialize to the case $p=2$, that is 
3 dimensions, and demonstrate the algebraic and geometrical 
structure of these theories. 

The main focus will be the deformation of the theory living on 
the boundary by the bulk theory. 
The couplings in the bulk can be viewed as elements of the cohomology 
of the deformation complex for the boundary theory. The path integral 
then calculates the corresponding deformation, as a straightforward 
generalization of deformation quantization. Indeed this was the basic 
strategy adopted in \cite{cafe}. 
The present paper will mainly deal with the semi-classical part of 
the quantization, that is it will only deal with the first order 
deformation. In a subsequent paper \cite{tomquant} we will discuss 
how to use path integral techniques to extent this to a full quantization, 
at least on a formal level. 

Perhaps the most interesting example, and indeed our primary 
motivation, is the open membrane coupling to a 
closed 3-form, which was called the open 2-brane in \cite{js}. 
This model has many interesting relations to both physics and mathematics. 
This model could be used to study the effect of the $C$-field
to the little string theory living on the 5-brane, 
\cite{bebe,kasa,iked,matshi,pio,bebe2,om}. In \cite{bebe,kasa} a 
constraint canonical quantization was used to study the model. 
This approach however is hard to extend to all orders; 
the method of BV quantization is much more suitable for this. 
Also the solution could become singular, as it involves the 
inversion of a 3-form. In this paper we will show that the 
topological open membrane coupling to the 3-form describes 
what is known as an exact Courant algebroid \cite{cour}. The classic 
Courant algebroid is based on the space $TM\oplus T^*M$, and 
was used to study general constraint quantization of gauged systems. 
It was shown that this Courant algebroid is deformed by a closed 3-form. 
Quantization of this object is still unsolved, but probably has a 
connection to gerbes. The deformed exact Courant algebroid controls a 
deformed version of quantization; the 3-form deforms a Poisson structure 
to a quasi-Poisson structure. In principle, the path integral for 
the open membrane model defines a formal quantization for this object. 

The simplest examples of our class of open membrane models 
are based on general quasi-Lie bialgebras, or Manin pairs $(\g,\g^*)$. 
These models are closely related to Chern-Simons theories. 
The relation between the topological open membrane and the quantization 
of the boundary string can be seen as a generalization of the relation 
between Chern-Simons and WZW models \cite{witpol}. The relation of the $G/G$ 
quotient WZW model and a double Chern-Simons \cite{gawed,falgaw} will 
explicitly appear as a special case of the topological membrane related 
to a particular quasi-Lie bialgebra. Quasi-Lie bialgebras are the 
infinitesimal objects related to (quasi-)Hopf algebras \cite{drin1}, 
a generalization of quantum groups. 
In our BV models the Yang-Baxter equation will be identified with 
part of the master equation, while scrooching/twisting of quasi-Lie 
bialgebras comes out simply as a canonical transformation. 
In a follow up paper \cite{tomquant} we will show that the boundary 
theory will have the structure of the corresponding quasi-Hopf algebras. 
Quantizability of the general Lie bialgebra 
was proven recently by Etinghof-Kazhdan \cite{etikaz}. 
The path integral of our model will give an alternative 
universal quantization formula for general quasi-Lie bialgebras. 

More generally we will find models based on Courant algebroids, 
which might also be called quasi-Lie bialgebroids. 
Algebroids combine the structure of tangent spaces and Lie algebras. 
Sections of the tangent bundle have a natural Lie bracket, 
which involves first order derivatives. Algebroids generalize this 
structure to more general fiber bundles. Lie bialgebroids 
can be described as dual pairs $(A,A^*)$ of Lie algebroids. 
The basic example is $A=TM$, which is equivalent to the exact 
Courant algebroid mentioned above \cite{royt}. Courant algebroids 
correspond to the generic topological open membrane. 
Recently Xu asked the question whether any Lie bialgebroid 
is quantizable \cite{xu}. A Lie bialgebroid is the 
geometrical structure underlying the classical dynamical Yang-Baxter 
equation. The corresponding quantum dynamical 
Yang-Baxter equation is relevant in quantizing Liouville theory, 
the Knizhnik-Zamolodchikov-Bernard equation, the Calogero-Moser 
model, and many related problems. Our approach gives a unified 
geometrical description of QDYB equations in terms of topological 
membrane theories. In particular, it gives a proposal 
for a formal universal quantization formula of Courant 
algebroids, based on 3-dimensional Feynman diagrams. 

The topological open membrane theories we will study give an interesting 
class of toy closed string field theories, \cite{zwie,wizwi,wit,kvz,ksv}
which in some cases can be solved exactly. In general closed string field 
theory has the structure of a $L_\infty$ algebra \cite{zwie}. 
In fact this $L_\infty$ algebra will play an important role in 
our discussion of the open membrane. It is this structure that will be 
deformed by the bulk deformations. 
Especially in the cases related to quasi-Lie bialgebras, the quasi-Hopf 
algebras will be constructed out of the closed string field theory. 
Other areas in physics where our model could be useful is the study of 
instanton effects in M-theory \cite{harmoor,mopesa} and 
the study of D-branes in the presence of a 3-form field strength 

This paper is organized as follows. 
In \secref{sec:bvtom} we introduce a convenient BV formulation 
in terms of superfields which allow us to give a simple 
geometric construction of topological membrane theories. 

In \secref{sec:master} we discuss the general algebraic 
structure of the master equation that follows from the semiclassical 
topological open membrane. 
In \secref{sec:alg} we discuss the semi-classical structure of 
the algebra of boundary operators for the open membrane. 

In \secref{sec:bialgebra} the simplest class of models related to 
quasi-Lie bialgebras are discussed in some detail. 

In \secref{sec:bialgebroid} we turn to topological open membranes 
based on exact Courant algebroid structures. These are related to 
membranes coupling to a closed 3-form. This is then generalized 
to more general Courant algebroids, combining the above situations 
of the tangent bundle and the quasi-Lie bialgebras. 
These models are the most general solutions of the master equation 
if one does not introduce negative ghost number superfields. 

In \secref{sec:courant} we review the mathematical structure of 
Courant algebroids, and show how our open membranes give rise to 
this structure. 

In \secref{sec:concl} we end with some conclusions and discussions 
on the results. 

While this paper was being finished, the paper \cite{iked2} appeared, 
which has some overlap with the present paper.

\section{BV Actions for Topological Open Membranes}
\label{sec:bvtom}

In this section we will develop a convenient description of a general class 
of BV actions for topological open membranes. 
We will only recall the main results of the detailed construction of \cite{js} 
relevant for the present paper.

\subsection{Superfields and BV Structure}

The theory of topological open $p$-branes developed in \cite{js}, 
specialized to the case $p=2$, involves an Euclidean open membrane 
living in a Euclidean target space $\CM$. The worldvolume theory of 
the membrane will be a topological theory, meaning that it does not 
depend on the worldvolume metric. The models studied in this paper 
will be manifestly independent of the metric, and be of BF type. 
The fields are differential forms, which have an action of the form 
\begin{equation}\label{bfaction}
  S_{BF} = \int_V \eta_{ij}B^i_{(2-p)} dA^j_{(p)} + interactions,
\end{equation}
where the index between brackets denotes the form degree and the 
interactions are formed by wedge products of the fields. 
Note that the form degree $p$ is at most 2. 
These theories have a lot of gauge symmetries which have to be gauge fixed. 
A general procedure to find a gauge fixed action is to use the BV 
formalism. For each of the fields $A^i$ and $B^i$, we need to introduce a 
whole set of ghost and antighost fields. The ghosts (and ghost-for-ghosts) 
for a $p$-form field $A^i_{(p)}$ will be corresponding lower degree fields. 
The antighosts are fields of all higher degree. It will be convenient 
to combine a field with all its ghosts and antighosts into a single 
superfield. These superfields can then be considered as maps between 
superspaces. Another advantage of using this superfield language is 
that it automatically takes care of some extra signs that 
are needed in the BV formulation. 

Quite generally, 
a topological field theory contains two operators of crucial importance: a BRST 
operator $\BQ$ and a fermionic operator $\BG_\mu$ transforming as a worldvolume 
1-form (the current of which is usually denoted $b$ in string theory). 
They satisfy the crucial anti-commutation relation $\{\BQ,\BG_\mu\}=\del_\mu$. 
Furthermore, there is a conserved ghost number charge called ghost, with 
$\BQ$ and $\BG$ having ghost numbers 1 and $-1$ respectively. 
Given any BRST closed worldvolume scalar operator $\CO$ we will define a set 
of descendants defined by $\CO^{(p+1)}=\BG\CO^{(p)}$, where $\CO^{(0)}=\CO$. 
These operators satisfy the descent equation $\BQ\CO^{(p+1)}=d\CO^{(p)}$, 
due to the anti-commutation relation above. As $\BG$ is a 1-form, 
the $p$th descendant $\CO^{(p)}$ will be a worldvolume $p$-form. 

Any physical field (of ghost number zero) will be the descendant of 
some scalar field $\phi^I$, generically a ghost. These scalars can be viewed as 
coordinates on a target superspace $\CM$. Equivalently, they can be seen 
as components of a map $\phi:V\to \CM$, where $V$ is the worldvolume. 
The coordinates on $V$ will be denoted $x^\mu$. As noted above, the $p$-form 
descendants of the coordinate fields can be combined into superfields 
which will be denoted $\Bphi^I$. 
For this purpose we introduce fermionic worldvolume coordinates $\theta^\mu$ 
of ghost degree $1$. Together the super coordinates $(x^\mu|\theta^\mu)$ 
can be viewed as coordinates on the superspace $\CV=\Pi TV$, where $\Pi$ 
denotes the shift of the degree by 1 (acting on the fiber). 
We will sometimes denote the supercoordinates collectively by $\Bx$. 
The supercoordinate fields are then functions of $(x^\mu|\theta^\mu)$ 
which can be expanded as 
\begin{equation}
  \Bphi^I(x,\theta) = \phi^I(x) +\theta^\mu\phi^{I(1)}_{\mu}(x) 
   + \frac{1}{2}\theta^\mu\theta^\nu\phi^{I(2)}_{\mu\nu}(x) 
   + \frac{1}{3!}\theta^\mu\theta^\nu\theta^\rho\phi^{I(3)}_{\mu\nu\rho}(x). 
\end{equation}
We treat the descendant components as separate fields. 
The descendant operator acts on superfields simply as 
$\BG_\mu=\frac{\del}{\del\theta^\mu}$. Combined together, the super coordinates 
can be viewed as a map between superspaces, $\Bphi:\CV\to\CM$. 
Note that if the superfield $\Bphi^I$ has ghost number $g$, the 
$p$th descendant will have ghost number $g-p$. 
The ghost number $g$ therefore equals the form degree of the 
physical field in the superfield.

Instead of starting with the BF theory and constructing a 
BV action we will start right away from the BV action. This will be 
a rather simple matter in the language of superfields. 
In order to define a BV structure for the membrane, the target space $\CM$ must 
be symplectic with symplectic form $\omega$. In this paper we will only 
consider constant $\omega$, though this restriction is not essential. 
This induces a symplectic form on the space of superfields by 
\begin{equation}
 \Bom_{BV} =  \int_\CV\Bphi^*\omega = \frac{1}{2}\int_\CV \omega_{IJ}\delta\Bphi^I\delta\Bphi^J,
\end{equation}
where $\delta$ denotes the De Rham differential on field space. 
Here the integral over $\CV$ involves integration over $x$ and $\theta$. 
It also defines a BV antibracket as the corresponding Poisson bracket, 
which we shall formally denote as follows, 
\begin{equation}
  \Bracket{\cdot,\cdot} = 
  \int_\CV \omega^{IJ}\frac{\del^R}{\del\Bphi^I}\wedge \frac{\del^L}{\del\Bphi^J}. 
\end{equation}
Here the $L$ ($R$) subscript indicates the left (right) derivative. These derivatives are 
functional derivatives with respect to the superfields $\Bphi^I$, defined in the usual way by 
\begin{equation}
  \frac{\del}{\del\eps}f(\Bphi+\eps\Bxi)\biggr|_{\eps=0} = \int_\CV\Bxi^I\frac{\del^Lf}{\del\Bphi^I}(\Bphi)
   = \int_\CV\frac{\del^Rf}{\del\Bphi^I}(\Bphi)\Bxi^I. 
\end{equation}
This BV bracket is derived from a BV operator, which is a second 
order differential operator formally given by 
\begin{equation}
  \BV = \frac{1}{2}\int_\CV \omega^{IJ}\frac{\del^2}{\del\Bphi^I\del\Bphi^J},
\end{equation}
where the derivatives are left-derivatives. 

The BV bracket $\Bracket{\cdot,\cdot}$ should have degree $1$, or 
equivalently the symplectic structure $\Bom_{BV}$ should have degree $-1$. 
Therefore, the symplectic structure $\omega$ on the target space must 
have degree 2, since the integration over $\CV$ has ghost degree $-3$. 
Hence we find that the target space $\CM$ is a symplectic supermanifold 
with a symplectic structure of degree 2. 

Let us recall some basic facts about BV algebras. 
The BV bracket is related to the BV operator by the relation 
\begin{equation}
  \Bracket{\alpha,\beta} = (-1)^{|\alpha|}\BV(\alpha\beta) 
-(-1)^{|\alpha|}\BV(\alpha)\beta  -\alpha\BV\beta. 
\end{equation}
The BV bracket is graded antisymmetric in the following shifted sense
\begin{equation}
\Bracket{\alpha,\beta} = -(-1)^{(|\alpha|+1)(|\beta|+1)}\Bracket{\beta,\alpha},
\end{equation}
and it satisfies the following graded Jacobi identity
\begin{equation}
\Bracket{\alpha,\Bracket{\beta,\gamma}} = \Bracket{\Bracket{\alpha,\beta},\gamma}
  +(-1)^{(|\alpha|+1)(|\beta|+1)} \Bracket{\beta,\Bracket{\alpha,\gamma} }.
\end{equation}

A BV action $S_{BV}$ determines a BRST operator by the relation $\BQ=\Bracket{S_{BV},\cdot}$. 
It squares to zero if the BV action satisfies the classical master equation 
$\Bracket{S_{BV},S_{BV}}=0$. Quantum mechanically this is not strictly necessary, 
but rather the BV action has to satisfy the quantum master equation 
$\BV S_{BV}+\frac{1}{2}\Bracket{S_{BV},S_{BV}}=0$. 
The Jacobi identity for the BV bracket implies the derivation condition for the 
BRST operator 
\begin{equation}
  \BQ\Bracket{\alpha,\beta} = \Bracket{\BQ\alpha,\beta} -(-1)^{|\alpha|}\Bracket{\alpha,\BQ\beta}. 
\end{equation}

Let us describe the structure of the target superspace. We will make use of 
the fact that for any supermanifold the the nonzero degrees form a fiber 
bundle over the degree zero submanifold, which we will denote $M$. 
In fact, if we denote by $\CM^p$ the submanifold of degree at most $p$, 
we find that $\CM^{p+1}$ is a fibration over $\CM^p$. For this paper 
we will assume that the target space is symplectic, or equivalently that the 
BV structure is nondegenerate. This can always be accomplished by adding 
extra fields. Furthermore we assume that all superfields will contain 
a physical (i.e. ghost degree zero) component. This reduces the degrees 
of the superfields, and thereby in the superspace $\CM$, to 0, 1, or 2. 
The degree 1 submanifold $\CM^1$ is a graded fiber bundle over $M=\CM^0$.  
As the BV structure is considered nondegenerate, there should be a natural 
(symmetric) pairing in the fiber. This implies that we can, at least 
locally, write the fiber bundle as $\CM^1=\A\oplus\A^*$. The fiber of 
degree 2 must be dual to the linearization of the degree 0 base. 
In other words it can be described by the fiber of the twisted 
cotangent bundle $T^*[2]M$.\footnote{In general $[p]$ denotes a shift 
of the (fiber) degree by $p$.} Combining this with the structure of the 
degree 1 fiber, we can describe the total target superspace as 
a twisted cotangent bundle $\CM=T^*[2]\A$. Here we used that the 
cotangent direction of the fiber is naturally the dual fiber, and the 
twist of the degree by 2 maps it degree back to 1. 

Locally the coordinates $\phi^I$ split into sets of conjugate coordinates 
$\phi^i$ on the base $\A$ and $\phi^+_i$ on the fiber. The shift implies 
that their degrees are related by $|\phi^+_i|=2-|\phi^i|$. 
The cotangent bundle comes with the canonical symplectic structure 
$\frac12\omega_{IJ}d\phi^Id\phi^J=d\phi^+_id\phi^i$. Due to the shift this 
has the required degree of 2. In the BV formulations, the conjugate 
superfields $\Bphi^+_i$ will contain the antifields of 
$\Bphi^i$, and vice versa.

\subsection{BV Action and BRST Operator}

The first part of the BV action will be given by the kinetic term, which 
in this paper will always be written in a first order form. Explicitly, 
our kinetic term will directly be determined by the BV structure and be 
given by\footnote{Here we assumed $\omega$ to be constant. In general 
the integrand is given in terms of a 1-form potential $\tau$ satisfying 
$d\tau=\omega$ as $\Bphi^*\tau=\tau_I(\Bphi)\Bd\Bphi^I$.}
\begin{equation}
  S_0 = \frac{1}{2}\int_\CV \omega_{IJ}\Bphi^I \Bd\Bphi^J, 
\end{equation} 
where $\Bd=\theta^\mu\frac{\del}{\del x^\mu}$ is the De Rham 
differential on the worldvolume in the superfield formalism. 
This action satisfies the classical master equation $\Bracket{S_0,S_0}=0$, and 
also the the quantum master equation, as $\BV S_{BV}=0$. This indeed 
has the BF form \eqref{bfaction}, whith the ``$A$'' and ``$B$'' fields 
residing in conjugate superfields with respect to the BV structure. 
The induced BV-BRST operator is given by $\BQ=\Bd$. This indeed 
satisfies the correct anticommutation relations with the operator $\BG$. 

The interaction terms in the action the membrane action will be given by a 
function of the superfields. The total bulk action will have the form 
\begin{equation}
  S= S_0+\int_\CV \Bga,
\end{equation}
where $\Bga(x,\theta)=(\Bphi^*\gamma)(x,\theta)=\gamma(\Bphi(x,\theta))$ 
for some function $\gamma\in C^\infty(\CM)$.\footnote{Here and in the following 
we will denote a pullback by the superfields by a boldface character.} 
We will require that $\gamma$ satisfies $\BV\Bga=0$, so that the classical 
master equation will imply the quantum master equation. The master 
equation then takes the form $\int \Bd\Bga+\frac{1}{2}\Bracket{\int\Bga,\int\Bga}=0$. 
If we can ignore boundary terms, the first term is a total derivative and 
therefore vanishes identically. Note that in order to get an action of ghost 
degree zero, $\gamma$ should be a function of degree 3. In the presence of the 
deformation $\gamma$, the BRST operator takes the form $\BQ=\Bd+\Bracket{\int\Bga,\cdot}$. 
The (classical) master equation is then indeed equivalent to $\BQ^2=0$. 
The anticommutation relation of the deformed BRST operator with 
$\BG$ is preserved by this deformation, due to the superfield structure. 
We can also add a boundary term of the form 
\begin{equation}
  \int_{\del\CV} \Bbe,
\end{equation}
where $\Bbe=\Bphi^*\beta$ for a function $\beta\in C^\infty(\CM)$ of degree 2, and 
$\del\CV=\Pi T(\del V)$ is the boundary of the super worldvolume.\footnote{Note 
that this is given by fixing both the even and odd normal coordinates.}

\section{Observables and the Master Equation}
\label{sec:master}

In this section we discuss the master equation of the class of topological 
open membranes introduced above. We formulate this in terms of a convenient 
algebraic framework related to the target space algebra.

\subsection{The Bulk Algebra}

First we discuss the precise relation between the field theory on the closed membrane 
to the algebra in the target space. In the rest of this section we discuss 
the generalization to open membranes. 

Observables for the bulk membrane can be found as functions of the superfields. 
They are therefore associated to functions on the target superspace. 
Let us denote this algebra of functions $\CA=C^\infty(\CM)$. 
The basic observable in the field theory on the membrane associated to $f\in\CA$ 
is the pullback to the super worldvolume $\CV$ of the membrane, 
$\Bf=\Bphi^* f$ where $\Bphi:\CV\to \CM$ is the map formed by the superfields. 
The BV symplectic structure on the superfields was inherited  by pullback 
of $\omega$ from the target space $\CM$. Let us denote the dual 
Poisson bracket on $\CA$ by $[\cdot,\cdot]$. This bracket is related to the 
BV bracket on field space by pullback, 
\begin{equation}
  \dint[_\CV]{\Bphi^*([f,g])} = \Bigl\Blbracket\dint[_\CV]\Bf,\dint[_\CV]\Bg\Bigr\Brbracket.
\end{equation}
The bracket $[\cdot,\cdot]$ in $\CA$ has degree $-2$, and therefore has the usual 
graded antisymmetry and Jacobi identity, rather than the shifted ones for the 
BV antibracket $\Bracket{\cdot,\cdot}$. 

Similarly, the BRST operator $\BQ$ in field space induces a nilpotent operator $Q$ 
on $\CA$. We have to be careful here, as the action involves a derivative on the 
worldvolume. And in our description using function on the target space, we did 
not included operators involving derivatives. To define $Q$ in the algebra $\CA$  
we will drop total derivatives over the worldvolume. For the closed membrane this 
will indeed be sufficient. Below we will be more careful 
about these contributions when we study the open membrane. 
With the above form of the action, we have 
\begin{equation}
  \BQ\dint[_\CV]\Bf=\dint[_\CV]{\Bd\Bf}+\Bigl\Blbracket\dint[_\CV]\Bga,\dint[_\CV]\Bf\Bigr\Brbracket.
\end{equation}
Dropping total derivatives, the operator $Q$ in the algebra is determined by 
the second term, and can be written $Qf=[\gamma,f]$. 

The algebraic structures on the target space are related to correlators in the field theory. 
For example, the bracket in the algebra $\CA=C^\infty(\CM)$ can be defined by the relation 
\begin{equation}
  \phi^*([f,g]) = \doint[_S]{(\phi^*f)^{(2)}}\phi^*g,
\end{equation}
where $S$ is a 2-cycle enclosing the insertion point of $g$. In terms of the superfields 
this can be written in the form $\Bphi^*([f,g]) = \doint[_{\Pi TS}]{\Bf}\Bg$. The integral 
over $\Pi TS$ includes in integral over two fermionic coordinates tangent to the cycle, 
and therefore picks out the first descendant when we specialize to the zeroth 
descendant component. 

The reason for the coincidence of the BV bracket with the above operator product 
is a result of the kinetic term, involving $\omega$ and $\Bd$. 
Using the (gauge fixed) propagator, this gives 
\begin{equation}
  \Vev{\doint[_S]{\phi^{I(2)}(x)}\phi^J(y)}
  \sim \omega^{IJ}\oint_S \frac{n_\mu(x-y)^\mu}{\|x-y\|^3} 
  \sim\omega^{IJ},
\end{equation}
where $n_\mu$ is the normal vector to the surface $S$. This correlation function 
is topological, and therefore only depends on the homology class of $S$. 

This is the structure of the closed membrane algebra. 
If we would introduce a boundary for the membrane, the above will still be 
valid when we assume that the observables $\Bf$ all vanish on the boundary, 
because then the total derivatives still vanish when integrated. 
This can actually be achieved by restriction on the algebra $\CA$. 
We will call this restricted bulk algebra $\CA_0$. This would describe 
the pure bulk theory. However, we are interested basically in what happens on 
the boundary.  We will now turn to the boundary algebra, which will be treated 
in a similar way.

\subsection{Including Boundary Terms}

The full target space is the superspace $\CM$. In the present paper, our main 
goal is the open membrane. Therefore, we have to specify boundary conditions. 
These will be determined by a choice of Lagrangian subspace $\CL\subset\CM$ 
(with respect to the BV structure). The boundary condition for the superfields is such 
that the boundary of the super-worldvolume $\del\CV=\Pi T(\del V)$ is mapped into 
this Lagrangian subspace $\CL$. The bulk operators were related to functions on the 
target space, giving the algebra $\CA=C^\infty(\CM)$. The Lagrangian condition 
ensures that the kinetic term $S_0$ satisfies the master equation $\Bracket{S_0,S_0}=0$, 
including the boundary term. 

As above, we consider a target space which is a twisted cotangent bundle, $\CM=T^*[2]\A$. 
A natural choice for the Lagrangian subspace $\CL$ is a section of this fiber bundle.  
In case $\CL$ is everywhere transverse to the fiber, we can canonically identify 
$\CL$ with the base $\A$.

The operators on the boundary can be interpreted as functions on the Lagrangian subspace 
$\CB=C^\infty(\CL)$. Given the Lagrangian subspace, we have a map $P_\CL:\CA\to \CB$ 
mapping functions on the total target space to functions on the Lagrangian subspace, 
defined by restriction. Note that the restricted bulk algebra mentioned above 
is given by $\CA_0=\ker P_\CL$. 

Taking into account the boundary term, 
the total BRST operator $\BQ$ acting on a bulk observable $\Bf=\Bphi^*f$ can be written 
\begin{equation}
\BQ\int_{\CV}\Bf = \BQ\int_{\CV}\Bphi^*f 
 = \int_{\CV}\Bphi^*(Qf)+\int_{\del\CV}\Bphi^*f.
\end{equation}
The first term indeed generates just the BRST operator in $\CA$, 
which we used above. In general, we have also the boundary 
term. We could set it to zero by demanding the extra condition $P_\CL f=f\bigr|_\CL=0$. 
Indeed, as $\Bphi$ restricted to the boundary maps into $\CL$, this gives a vanishing 
boundary term. These functions represent the pure bulk operators. More generally, 
we incorporate the boundary terms into our description by 
extending the space of operators to $\bar \CA= \CA\oplus \CB$ including both 
the bulk and the boundary deformations. Elements are pairs $f\oplus g\in \CA\oplus \CB$, 
for which we define the ($\Bphi$-dependent) formal integral 
\begin{equation}
\int f\oplus g \equiv \int_{\CV}\Bf +\int_{\del\CV}\Bg. 
\end{equation}
We can interpret the 
restriction map $P_\CL$ as an off-diagonal map in this extended algebra 
$P_\CL:f\oplus g\mapsto 0\oplus P_\CL f$. 
Note that this operation trivially squares to 0. It is in fact the unperturbed 
BRST operator, for $\gamma=0$. 
With these notations, we can write the above identity --- also including a 
boundary term --- in the form $\BQ\int f\oplus g=\int Qf\oplus (P_\CL f-Q_\CL g)$, 
where $Q_\CL$ denotes the restriction of $Q$ to the boundary. Here the relative 
minus sign in front of $Q_\CL$ is due to the fact that $\int_{\del\CV}$ has 
degree $-2$ (or equivalently, it involves a degree one delta-function on the boundary). 
This leads to a BRST operator on the extended operator 
space $\bar \CA$ having the block form 
\begin{equation}
\bar Q=\pmatrix{Q & 0\cr P_\CL &-Q_\CL}:\,\CA\oplus \CB\to \CA\oplus \CB.
\end{equation}
The relation $Q_\CL P_\CL=P_\CL Q$ ensures that $\bar Q^2=0$. 

We also need to know how the bracket extends to the total space $\bar \CA$. 
The bracket will be zero when restricted to the boundary, due to the Lagrangian 
boundary condition. So we only need to give the prescription for the bracket acting 
between $\CA$ and $\CB$. To find an expression for this we will use the derivation 
condition of the unperturbed BRST operator $P_\CL$, 
\begin{equation}
  P_\CL[\alpha,\beta] = [P_\CL\alpha,\beta] +(-1)^{|\alpha|}[\alpha,P_\CL\beta], 
\end{equation}
which is a consequence of the corresponding identity in field space. 
To give a more explicit description, we will need an explicit embedding 
$i_\CL:\CB\to \CA$, satisfying $P_\CL\circ i_\CL=1_\CB$. For 
$\alpha=f\oplus0\in\CA_0$ and $\beta=i_Lg\oplus0$ the above implies
\begin{equation}
  [f\oplus 0,0\oplus g] = 0 \oplus (-1)^{|f|}P_\CL[f,i_\CL g].
\end{equation}
This will be independent of the choice of embedding $i_\CL$ due to the above identity. 
For $P_\CL f\neq 0$, the simplified description in terms of the algebra 
will not be sufficient anymore. We will however not need this generalization. 
Of course, this result can also be derived from the BV bracket on field space.

\subsection{Deformations, BRST Cohomology and Canonical Transformations}

Infinitesimal deformations of the action are controlled by the BRST 
cohomology. This should be the cohomology for the total BRST operator 
$\bar Q$. The total space $\bar\CA$ can be viewed as the total 
complex of a double complex, with differentials $P_\CL$ and $Q$. 
The total cohomology can be calculated using spectral sequence techniques. 
In the following calculation we will assume that $Q_\CL=0$ for simplicity, 
although one can easily generalize. 

We decompose $\bar Q =Q+P_\CL$, and first take cohomology with respect to $Q$. 
The first term in the spectral sequence is then $E_1=H_Q(\CA)\oplus \CB$, 
as $Q$ acts only on $\CA$. The term $E_1$ has differential induced by $P_\CL$. 
We denote by $P_\CL':H_Q(\CA)\to \CB$ the induced projection $P_\CL$ reduced to $H_Q(\CA)$. 
Note that this is well defined, as $P_\CL=0$ on $\im Q$ by our assumption. 
Taking its cohomology restricts the bulk term to elements in the kernel of $P_\CL'$. 
In other words, the bulk deformations are $Q$-cohomology classes 
vanishing on the boundary. 
The boundary term is defined up to the image of $P_\CL'$. 
The spectral sequence terminates at the second term because there is no room 
for higher differentials. 
We conclude $H_{\bar Q}(\bar \CA)\cong E_2\cong \ker P_\CL'\oplus (\CB/\im P_\CL')$. 
For nonzero $Q_\CL$, we should have replaced $\CB$ by $H_{Q_\CL}(\CB)$. 

An alternative way to calculate the cohomology is to start the spectral sequence 
with $P_\CL$. Then the first term is given by $E_1=H_{P_\CL}(\bar \CA)=\CA_0\oplus 0$, 
as $P_\CL$ is surjective. 
Denoting $Q'=Q\bigr|_{\CA_0}$, we have $E_2=H_{Q'}(\CA_0)\oplus 0$. 
The spectral sequence terminates at the second term, as $E_1$ is concentrated 
in a single degree (in the $P_\CL$ direction). Therefore 
$H_{\bar Q}(\bar \CA)\cong E_2\cong H_{Q'}(\ker P_\CL)$. 

The two answers do not look the same. For example, the first one contains 
boundary terms, while the second has only bulk deformations. 
The two results are however equivalent. We will see below how boundary 
deformations can be turned into bulk terms in vice versa by canonical transformations. 

The BRST cohomology is closely related to canonical transformations in the BV theory. 
For any function $\beta\in\CA$ let us define the operator $\delta_\beta=[\cdot,\beta]$. 
In the following we will mainly use $\beta$ of degree 2. It is basically the Hamiltonian 
vector field with respect to the symplectic structure. Similarly, on superfield space 
we define the operator $\Bde_\Bbe=\Bracket{\cdot,\int\Bbe}$. This operator is the 
generator of a canonical transformation. The relation between the BRST cohomology 
and a canonical transformation is based on the following relation
\begin{equation}
\e^{t\Bde_\Bbe}S = S+t\BQ\int\Bbe+\CO(t^2). 
\end{equation}
In other words, to first order in $t$ a canonical transformation shifts the action 
by a BRST exact term. The first order shift of the action by a BRST exact term usually 
does not produce a solution of the master equation. It can however be turned into 
a solution of the master equation by adding higher order corrections, which 
are generated by the full canonical transformation. A canonical transformation is 
a true symmetry of the theory, while the BRST exact terms only give an approximation. 

In terms of the algebraic language we have developed above, and in case we can 
ignore boundary terms, the above can be reduced to the algebra $\CA$, 
\begin{equation}
\e^{t\delta_\beta}\gamma = \gamma+tQ\beta+\CO(t^2).
\end{equation}
Even if there are boundary terms, the term $\e^{\delta_\beta}\gamma$ is still 
a solution of the bulk master equation when $\gamma$ is. It is a solution to the full 
master equation if in addition the boundary master equation 
$P_\CL\bigl(\e^{\delta_\beta}\gamma\bigr)=0$ is satisfied. The solution however 
is not necessarily equivalent to $\gamma$, as the canonical transformation can 
produce boundary terms, which we have here ignored.

\subsection{Boundary Deformations}

We next consider the case where the boundary term does not vanish. 
Actually, we can use what we have found above for the case where there is no 
boundary term. 

First, we have to be careful about the kinetic term in the action. In 
general, we write the full action as $S_0+\Gamma$, where $S_0$ is the kinetic term 
and $\Gamma$ is assumed to be the integral of the pull-back of a function $\gamma$ 
on $\CM$. Furthermore, we will deform the action by an extra boundary 
term, which is the integral of a pullback from $\A$. Note that for $\gamma=0$ 
we have $Q=Q_\CL=0$. 

First we note that 
\begin{equation}
  \Bde_\Bbe S_0 = \Bigl\Blbracket S_0,\dint[_\CV]\Bbe\Bigr\Brbracket
   = \int_{\CV}\Bd\Bbe = \int_{\del \CV}\Bbe.
\end{equation}
To be able to describe this in terms of the algebra $\CA\oplus \CB$, 
we adjoin to the bulk algebra $\CA$ a 
formal element $\tau$ corresponding to $S_0$, i.e.\ formally $S_0=\int\tau$, and satisfies 
\begin{equation}
  \delta_\beta \tau = 0\oplus P_\CL\beta
\end{equation}
for any $\beta$. Then we have
\begin{equation}
  \e^{\Bde_\Bbe}\int(\tau+\gamma)\oplus 0 
  = \int(\tau+\e^{\delta_\beta}\gamma)\oplus 
    \biggl(\sum_{n\geq 1}\frac{1}{n!}(\delta_\beta)^{n-1} P_\CL\beta\biggr).
\end{equation}
We assume that $[\beta,\beta]=0$, so that only the $n=1$ term survives in the boundary term. 
An important case where this is satisfied is when $\beta\in i_\CL(\CB)$. 
If we ignore the boundary term, we find what we 
used before: the canonical transformation of the kinetic term is a total derivative, 
and therefore trivial, so we only transform the bulk deformation $\gamma$. We know that 
the pure bulk term $S_0+\Gamma_{\gamma,\beta}=S_0+\e^{-\Bde_\Bbe}\dint\Bga$ is a solution 
to the full master equation if $\gamma$ is a solution of the bulk 
master equation, i.e. $[\gamma,\gamma]=0$, and the boundary term vanishes, 
$P_\CL(\e^{-\delta_\beta}\gamma)=0$.  However, if $P_\CL\beta\neq 0$, 
this solution is not equivalent to the solution $S_0+\int\gamma$. 
In fact, we have 
\begin{equation}
\e^{\Bde_\Bbe}(S_0+\Gamma_{\gamma,\beta})
  = \e^{\delta_\beta}\int(\tau+\e^{-\delta_\beta}\gamma) 
  = \int(\tau+\gamma)\oplus P_\CL\beta = S_0+\int \gamma\oplus P_\CL\beta,
\end{equation}
where we assumed that $[\beta,\beta]=0$ to prevent higher order terms in 
the boundary term. As this includes all contributions of the canonical 
transformation, it should be equivalent to the full action $S+\Gamma_{\gamma,\beta}$.  
So we have actually written the deformation 
using $\beta$ in terms of a boundary term. Therefore, if we can solve our constraint 
of vanishing field strength, we can add a boundary term to cancel the boundary term 
in the master equation. So although the action looks simple, the BV master equation 
is much more nontrivial due to the boundary term. In general, it can be found 
by writing the terms again as superfields, and the boundary term as a bulk term 
using $\Bd$, writing the master equation for the bulk and writing total 
derivatives again as boundary terms. 
It has in general two components: a bulk and a boundary term, given by 
\begin{equation}
  Q\gamma+\frac{1}{2}[\gamma,\gamma] = 0,\qquad
  P_\CL\gamma = 0. 
\end{equation}
At first sight, this seems to be the master equation for $\beta=0$, rather 
than the one for nonzero boundary term to which it is supposed to be equivalent. 
We have to be very careful however with the boundary 
condition for the fields, as they are different in both cases. 
Assume that before the canonical transformation we had a boundary condition 
$\psi^i|_{\del V}=0$. After the canonical transformation, we have changed the fields, 
which means that in the new variables the boundary condition becomes
\begin{equation}
  \e^{-\delta_\beta}\psi^i\Bigr|_{\del V} = (\psi^i-[\psi^i,\beta])\Bigr|_{\del V} = 0. 
\end{equation}
Thus can also be found by realizing that variation with respect to $\chi_i$ 
has a boundary term $\delta\chi_i\Bigl(\psi^i-\frac{\del\beta}{\del\chi_i}\Bigr)$. 
As $\delta\chi_i$ is arbitrary on the boundary, this requires the above boundary 
condition for $\psi^i$. 

This implies that the projector $P_\CL$ has changed due 
to the presence of the boundary term $\beta$. To see how, 
let us call the original projector $P_\CL^0$, and the projector in the presence 
of a boundary term $P_\CL^\beta$. These two operators are then related by 
a canonical transformation as 
\begin{equation}
  P_\CL^{\beta} = \e^{\delta_\beta}\circ P_\CL^0\circ\e^{-\delta_\beta}. 
\end{equation}
The boundary master equation has to be interpreted as $P_\CL^\beta\gamma=0$. 
This is indeed the same as the original constraint 
$P_\CL^0(\e^{-\delta_\beta}\gamma)=0$ we found for the equivalent pure bulk action. 
Expanding the exponential, this can be written in the form 
\begin{equation}\label{bdymaster}
  \sum_{n\geq0}\frac{1}{n!}P_\CL^0(-\delta_\beta)^n\gamma 
  = P_\CL^0\gamma - P_\CL^0[\gamma,\beta] + \frac{1}{2}P_\CL^0 [[\gamma,\beta],\beta] + \cdots =0.
\end{equation}
Later, we will give an interpretation of the various terms in this equation.

\section{The Algebraic Structure of Open Membranes}
\label{sec:alg}

We will now discuss the general structure of the deformed boundary algebra that arises 
as sketched above. We will see that in general there is a structure of $L_\infty$ algebra, 
which arises in a way we call a \emph{derived} $L_\infty$ algebra, 
generalizing the notion of derived bracket.

\subsection{Correlators and the Boundary Algebra}

Let us first discuss the correlation functions of boundary operators in the open 
membrane theory in the presence of a nontrivial bulk term $\gamma$. As we discussed 
the basic boundary observables are determined by functions on the Lagrangian $\CL\subset\CM$. 

First we write the action as the sum of a kinetic term and an interaction term, 
$S=S_0+S_{int}$, where we took $S_{int}=\int\Bga$. 
Using a Gaussian integral in the path integral, we can write the correlation functions as 
\begin{equation}
  \Vev{\prod_a\CO_a} = 
 \dint{\CD\Bphi}\e^{i\Pi\bigl[\frac{\del}{\del\Bphi}\bigr]}\biggl(\e^{iS_{int}[\Bphi]}\prod_a\CO_a\biggr).
\end{equation}
The propagator in the above expression, seen as a bidifferential operator, 
can be written in the form 
\begin{equation}
\Pi\biggl[\frac{\del}{\del\Bphi}\biggr] = 
\dint[_\CV]{d\Bx}\dint[_\CV]{d\By}\Pi(\Bx,\By)
\omega^{IJ}\frac{\del}{\del\Bphi^I(\Bx)}\frac{\del}{\del\Bphi^J(\By)}.
\end{equation}
Here $\Pi(\Bx,\By)$ is the integral kernel for the inverse kinetic operator $\Bd\inv$ 
(after gauge fixing). 
We recognize in this expression the BV bracket structure. Because of this we will see 
that we can effectively describe the algebraic structure on the boundary operators 
in terms of the original BV bracket. 

The boundary theory is basically a topological closed string theory. As discussed in 
\cite{homa2}, one of the essential operations in the algebra of observables is based 
on the bracket determined by the contour integral of one operator around another, 
\begin{equation}
  \{f,g\} = \doint[_C]{f^{(1)}}g,
\end{equation}
where $C$ is a 1-cycle enclosing the insertion point of $g$. Interpreting the 
1-form $f^{(1)}$ as a worldsheet current, this is actually the action of the current 
on a scalar operator. This bracket determines the current algebra in the string theory. 
For example, a Ward identity implies that the supercommutator 
$[\doint{f^{(1)}},\doint{g^{(1)}}]=\doint{\{f,g\}^{(1)}}$.

The bracket $\{\cdot,\cdot\}$ introduced above is an antibracket of degree $-1$. 
Therefore, the graded antisymmetry and Jacobi identity are similar to those of the 
BV antibracket $\Bracket{\cdot,\cdot}$ on field space. This is part of the reason that 
the closed string algebra forms has the on-shell structure of a BV algebra \cite{zwie}.

This bracket can again conveniently be written in terms of the superfields. We 
introduce the super 1-cycle $\CC=\Pi TC$; the integration over the fermionic 
coordinates picks up the first descendant in the tangent direction. More 
precisely, we can define the operation in terms of the correlation function 
\begin{equation}
  \Vev{\Bde_{\phi_0}\doint[_\CC]\Bf\Bg},
\end{equation}
where all the operators are put on the boundary and $\delta_{\phi_0}$ is a delta function 
fixing the scalar fields to a fixed value $\phi_0$ consistent with the boundary condition. 
After contractions, and using the expression for the propagator above, the lowest 
order term can be written 
\begin{equation}
  \dint[_\CV]{d\Bz}\doint[_\CC]{d\By}\Pi(\Bz,\By)\Pi(\Bz,\Bx)
  \dint{d\phi}\delta(\phi-\phi_0)\omega^{KL}\omega^{IJ}\frac{\del^2\gamma}{\del\phi^K\del\phi^I}\frac{\del f}{\del\phi^J}\frac{\del g}{\del\phi^L}. 
\end{equation}
This is just the Feynman integral corresponding to a 2-legged tree-level diagram. 
The integral is a universal factor, that does no longer depend on the precise choice 
of operators. The dependence on the functions $f$ and $g$, and therefore the choice 
of boundary observables is expressed in terms of differential operators acting on 
these functions. To see that this is nontrivial, one should check that the integral 
indeed is a number different from zero. That this is indeed the case will be shown 
elsewhere \cite{tombv}. In terms of the the boundary algebra of functions 
$\CB=C^\infty(\CL)$, the bracket can now be written 
(after a proper normalization and including signs)
\begin{equation}\label{bdybracket}
  \{f,g\} = (-1)^{|f|+1}P_\CL[[\gamma,f],g] +(-1)^{|f|(|g|+1)}  P_\CL[[\gamma,g],f].
\end{equation}
Here the $P_\CL$ results from the projection on the outgoing state $\Bde_{\phi_0}$, 
or the delta-function in the zero-mode integral over $\phi$. 
More precisely, we should interpret the boundary operators like $f$ as 
embedded in the algebra $\CA$; so we should write $i_\CL f$. 

In the above form of the bracket, the reader can readily recognize the structure of 
a term in the boundary master equation we met before. This is no coincidence, and is 
a direct consequence of the equivalence between the deformed theories with and without 
a boundary term. 

More general correlation functions can be found by introducing more integrated 
operators. The operator products related to the brackets in the boundary string 
are given by the operator equation 
\begin{equation}\label{brackop}
  \{\Bf_{1},\cdots,\Bf_n\} = (-1)^{|f_1|+|f_2|+\ldots+|f_{n-2}|}
  \dint[_{\del\CV}]{\Bf_1}\cdots\dint[_{\del\CV}]{\Bf_{n-2}}\doint[_\CC]{\Bf_{n-1}} \Bf_n
  + perms,
\end{equation}
where $\CC=\Pi TC$ is a super 1-cycle in the boundary. In terms of the components 
of the superfields, this can be written 
\begin{equation}
  \{f_1,\cdots,f_n\} = (-1)^{|f_1|+|f_2|+\ldots+|f_{n-2}|}
  \dint[_{\del V}]{f_1^{(2)}}\cdots\dint[_{\del V}]{f_{n-2}^{(2)}}\doint[_C]{f_{n-1}^{(1)}} f_n
  + perms. 
\end{equation}
The corresponding correlation functions are topological. The relevance of these 
operations and the relation to the $L_\infty$ structure was explained in \cite{homa2}. 
They can be interpreted as the structure constant for the bosonic closed 
string field theory \cite{zwie} of the corresponding boundary string. 

The semiclassical approximation to these brackets are calculated analogously 
to that of the bracket, involving the various contractions. Due to the form 
of the propagator the brackets in the boundary algebra are 
induced by the BV bracket in the bulk and the bulk term in the action. 
In fact, the form \eqref{bdybracket} is almost that of the well known mathematical 
notion of a \emph{derived bracket}.

\subsection{Derived $L_\infty$ Algebra}

We can express the above results of the boundary brackets in our algebraic 
language in terms of the basic structure, the bracket and the BRST operator. 
This gives rise to a generalization of so-called derived brackets. 

Let us assume a graded differential Lie algebra $\CA$. This means that it is provided 
with a Lie-bracket $[\cdot,\cdot]$ of degree $p$ and a derivation $d$ 
of this bracket which squares to zero. 
Then we can define the \emph{derived bracket} $\circ$, of degree $p+1$, by 
\begin{equation}
  f\circ g = (-1)^{|f|+1}[df,g].
\end{equation}
In general this derived bracket is not skew symmetric. It satisfies a close 
analog of the Jacobi identity, making $\CA$ into a Loday algebra. 
The differential $d$ is also a derivation of the derived bracket. 
We could have also considered the skew-symmetrization of $\circ$, 
which is sometimes called the derived bracket. This bracket 
will in general not satisfy the Jacobi identity. The derived bracket becomes important 
when we study an abelian subalgebra $\CB$ of $\CA$ (with respect to $[\cdot,\cdot]$). 
It can be shown that restricted to $\CB$ the derived bracket is actually skew-symmetric. 
When $\CB$ in addition is closed with respect to $d$ and $\circ$, 
it is a graded differential Lie algebra itself, with a bracket of degree $p+1$. 

A well known example of a derived bracket is a Poisson bracket. Consider a manifold 
$M$ and the algebra $\CA=C^\infty(\Pi T^*M)=\Gamma(\ext[*]TM)$ is the algebra of 
multivector fields. This is naturally provided with the Schouten-Nijenhuis bracket 
$[\cdot,\cdot]$ (the generalization to multivector fields of the Lie bracket). 
Now we choose a bivector $\pi$ satisfying $[\pi,\pi]=0$ (Poisson structure), 
and consider the derivation $d_\pi=[\pi,\cdot]$. This makes $\CA$ into a differential 
Lie algebra. The algebra $\CB=C^\infty(M)$ of functions is an abelian subalgebra stable 
with respect to $d_\pi$. The derived bracket on $\CB$, given by 
$\{f,g\}=(-1)^{|f|+1}[[\pi,f],g]$, is precisely the Poisson bracket generated by $\pi$. 
This example is actually the analogous boundary bracket for the open string of the 
Poisson-sigma model \cite{cafe}. 

We now observe that the tree level result for the bracket 
\eqref{bdybracket} in the boundary algebra has the form of 
the skew-symmetrization of a derived bracket, with $d=[\gamma,\cdot]=Q$, 
apart from the projection $P_\CL$. This projection had to be inserted 
because the boundary algebra $\CB$ is not closed under the 
derived bracket. Indeed, this is a natural extension of an 
induced derived bracket. More generally at tree level, 
the differential (BRST operator), bracket and trilinear bracket 
on the boundary algebra $\CB$ are given by the expressions 
\begin{eqnarray}\label{derlinfty}
  Q_\CL f &=& P_\CL[\gamma,i_\CL f],\nonumber\\
  \{f,g\} &=& (-1)^{|f|+1}P_\CL[[\gamma,i_\CL f],i_\CL g] \pm perms.,\\
  \{f,g,h\} &=& (-1)^{2|f|+|g|+3}P_\CL[[[\gamma,i_\CL f],i_\CL g],i_\CL h] \pm perms.\nonumber
\end{eqnarray}
One could go on, but at least semi-classically, the higher brackets all 
vanish in the theories we study, due to the degree. This is an 
obvious generalization of the notion of derived bracket to higher brackets. 
Note that the induced derived bracket on $\CB$ with the projection 
$P_\CL$ is not a Lie bracket in general. However as it turns out 
they do satisfy the relations of an $L_\infty$ algebra or homotopy 
Lie algebra. We will therefore call this a \emph{derived $L_\infty$ algebra}. 

We can now recognize the boundary master equation \eqref{bdymaster} 
as the Maurer-Cartan equation $Q_\CL\beta+\{\beta,\beta\}+\cdots=0$ 
of this derived $L_\infty$ algebra. The bilinear boundary bracket 
should be interpreted as the BV bracket of the boundary string. 
We note that the BV algebra in the boundary string is of a more 
general homotopy type, including higher brackets. Such generalizations 
of the BV algebra appeared in the context of BV quantization 
in \cite{batmar2}, were they were called quantum antibrackets.

\subsection{Path Integral Quantization and Deformation Theory}

Now that we have described the semiclassical deformation structure 
of our model, let us shortly discuss how to pass to the quantization. 
This will be a generalization of the problem of deformation quantization 
for the associative algebra of functions. 

Topological field theories in $d$ dimensions are closely 
related to $d$-algebras. Indeed, $d$-algebras can be defined in terms 
of the homology of configuration spaces of punctured $d$-dimensional discs 
\cite{getzjon2,kon1}. A $1$-algebra is simply an associative algebra, while 
for $d\geq2$, a $d$-algebra in general is a (super)commutative associative 
algebra with a twisted Lie-bracket of degree $d-1$ \cite{kon1}.
Particularly important examples of (super)commutative 
algebras are provided by the algebra of functions $\CB=C^\infty(\A)$ 
on some (super)manifold $\A$. They become $d$-algebras when provided 
with a (possibly zero) twisted Lie bracket of degree $d-1$. 
The BV sigma model canonically associates a topological 
membrane theory to any such $2$-algebra. 
Our quantization can be considered as expressing the generalized 
Deligne conjecture, which states that the deformation of a 
$d$-algebra is a $(d+1)$-algebra, see for example \cite{tam2}. 
We interpret this by saying that the $(d+1)$-dimensional 
topological field theory deforms the $d$-dimensional topological 
field theory on the boundary. The Hochschild cohomology---closely related 
to the deformation complex---of $\CB=C^\infty(\A)$ as a $d$-algebra 
is given by the algebra of functions on the twisted cotangent space, 
$H\!H^*(\CB)=C^\infty(T^*[d]\A)$, c.f. \cite{kon2}. This is naturally 
reflected in our BV sigma models, where the target superspace of the 
bulk membrane has the form $\CM=T^*[2]\A$, with the boundary string living in $\A$. 

The objective of the quantization program will be to construct a map 
from the Hochschild cohomology $\CA=H\!H^*(\CB)$ to the Hochschild complex of the 2-algebra, 
\begin{equation}
  \mathcal{Q}: C^\infty(\CM) \to C^*(\CB,\CB). 
\end{equation}
This map should be intertwining, at least up to a quasi-isomorphism. 
It then gives a formality of the complex as a $G_\infty$ 
or homotopy Gerstenhaber algebra. 
The map will be constructed using the path integral of the 
topological open membrane corresponding to the $G$ algebra $\CB$. 
For this we also need an inner product, which is provided by the 2-point 
function. For $\gamma\in C^\infty(\CM)$ the proposed quantization map 
is given by 
\begin{equation}
  \Vev{f_0,\mathcal{Q}_C(\gamma)(f_1,\cdots,f_n)} 
  = \dint{\CD\Bphi} \e^{\frac{1}{\hbar}(S_0+S_\gamma)}\CO_C(f_0,\cdots,f_n),
\end{equation}
where the deformed action is given by $S_\gamma = \int_\CV\Bga$. 
$\CO_C$ is a boundary observable composed out of the operators 
$f_i\in\CB$ and depends on an extra label, which runs over chains in the 
configuration space of the $n+1$ insertion points. They run over 
the labels of the maps defining the $G_\infty$ structure. So any 
$\gamma\in C^\infty(\CM)$ gives rise to a whole set of multilinear maps 
$\mathcal{Q}_C(\gamma)\in C^{n_C}(\CB,\CB)=\Hom(\CB^{\otimes n_C},\CB)$. 
The brackets of the $L_\infty$ algebra correspond to particular 
examples of the boundary observables \eqref{brackop}. 
Indeed, this is precisely a reduction in the topological context 
for the $L_\infty$ algebra in string field theory \cite{senzwie}. 
We can find a quantization of the full $G_\infty$ algebra by 
considering more general observables. For example, we should include 
an observable $\CO_{A,2}(f_1,f_2) = \Bf_1\Bf_2$ for the product, 
at least to lowest order. A more detailed discussion will be given 
in a forthcoming paper \cite{tomquant}. 

The path integral can be perturbatively expanded as a sum over 
Feynman diagrams, each corresponding to a particular term in the 
expansion of the products, and a universal weight given by an 
integral involving CS propagators $\Bd\inv$. As the above quantization 
map is intertwining, a solution to the master equation in $C^\infty(\CM)$ 
is mapped to a deformed $G_\infty$ structure in the complex.

\section{Topological Membranes from Quasi-Lie Bialgebras}
\label{sec:bialgebra}

In this section we discuss a particular class of open membrane models 
based on a purely fermionic target space. The models, which are of a 
Chern-Simons type, have a semi-classical structure of a quasi-Lie bialgebra 
or Manin pair. This allows us to interpret the open membrane model as 
a quantization of these mathematical objects, which are quasi-Hopf algebras or 
quantum groups.

\subsection{Quasi-Lie Bialgebras and Open Membranes}

Above we have described a construction for topological membranes based on 
a twisted cotangent bundle of a supermanifold. As the symplectic structure 
should have degree two, the simplest way to get this is to take a manifold 
which is completely of degree one. This is the class of models we will study 
in this section. More explicitly, the target space will be given by 
$\CM=T^*[2](\Pi\g)=\Pi(\g\oplus\g^*)$, where $\g$ is any vector space and 
$\g^*$ its dual. In other words, the base space is the graded space $\A=\Pi\g$. 
Hence we will initially take the Lagrangian to be the zero section, $\CL=\Pi\g$. 
We will choose flat coordinates $\chi^i$ and $\psi_i$, on the base and the 
fiber respectively. The Lagrangian submanifold $\CL$ is then given by the 
equations $\psi_i=0$. This gives two superfields of ghost degree one, 
\begin{equation}
  \Bps_i = \psi_i+\theta B_i+\cdots,\qquad
  \Bch^i = \chi^i +\theta A^i+\cdots.
\end{equation}
The action for this model following the general description takes the form 
\begin{equation}
  S = \int_\CV\Bigl(\Bps_i \Bd\Bch^i + \gamma(\Bps,\Bch)\Bigr). 
\end{equation}
The function $\gamma$ is cubic, and has the general form 
\begin{equation}
  \gamma = \frac{1}{2}c^i_{jk}\psi_i\chi^j\chi^k 
   + \frac{1}{2}f^{ij}_k\psi_i\psi_j\chi^k 
   + \frac{1}{3!} \varphi^{ijk}\psi_i\psi_j\psi_k.
\end{equation}
The $\chi\chi\chi$ term has to vanish in order for the condition $P_\CL\gamma=0$ to be satisfied. 
The condition $\triangle\gamma=0$ implies the vanishing of the traces 
$c^i_{ij}=0=f^{ij}_i$. 
The master equation $[\gamma,\gamma]=0$ is equivalent to the 4 relations 
\begin{equation}
\renewcommand{\arraystretch}{1.2}
\begin{array}{r@{\;}c@{\;}l@{\qquad}r@{\;}c@{\;}l@{\;}}
  c^m_{[ij}c^l_{k]m} &=& 0,&
  \frac{1}{2}f^{[ij}_mf^{k]m}_l+c^{[i}_{lm}\varphi^{jk]m} &=& 0,\\
  c_{m[i}^{[k}f_{j]}^{l]m} &=& 0,&
  f^{[ij}_m\varphi^{kl]m} &=& 0.
\end{array}  
\end{equation}
The first condition implies that $c_{ij}^k$ are the structure constants of a 
Lie-algebra, based on the vector space $\g$. 
When the $\varphi^{ijk}$ vanish, we see from the second equation that the $f^{ij}_k$ 
also are the structure constants of a Lie-algebra. In other words 
$\g^*$ is also a Lie-algebra in this case. Note that always the 
total space $\g\oplus\g^*$ has the structure of a Lie algebra.\footnote{We denote 
here the total space by $\g\oplus\g^*$, which is only true as a vector space. 
The reader should be aware that as a Lie-algebra it is not simply a direct sum.} 

For more concreteness, let us introducing a basis $e_i$ for $\g$ and a 
dual basis $e^i$ for $\g^*$. With respect to this basis, the Lie-bracket 
on $\g\oplus\g^*$ can be written as 
\begin{eqnarray}
  [e_i,e_j] &=& c_{ij}^ke_k, \nonumber\\{}
  [e_i,e^j] &=& f_{i}^{jk}e_k - c_{ik}^je^k, \\{}
  [e^i,e^j] &=& f_k^{ij}e^k + \varphi^{ijk}e_k. \nonumber
\end{eqnarray}

When $\varphi^{ijk}$ vanishes, we see that both $\g$ and $\g^*$ have the structure 
of a Lie algebra, with some extra compatibility condition between the two structures. 
One also calls the triple $(\g\oplus\g^*,\g,\g^*)$ a \emph{Manin triple} in this case. 
It consists of a Lie algebra $\g\oplus\g^*$ with an invariant nondegenerate inner 
product and two isotropic Lie subalgebras. The structure constants $f_i^{jk}$ 
can also be interpreted as a so-called cocommutator, a map $\delta:\g\to\ext[2]\g$, 
given by  
\begin{equation}
  \delta(e_i) = f_i^{jk}e_j\wedge e_k.
\end{equation}
The above conditions say that $\delta$ squares to zero and is a cocycle. 
The Lie algebra $\g$ with the cocommutator $\delta$ is called a \emph{Lie bialgebra}. 
Note that this notion is dual, as also $\g^*$ is a Lie bialgebra. 
The commutator of $\g^*$ is dual to the cocommutator of $\g$, and vice versa. 
When only $c_{ij}^k$ is nonzero, we find the canonical Lie bialgebra structure, 
consisting of the Lie bracket $\ext[2]\g\to\g$ and the adjoint action of 
$\g$ on its dual, $\g\otimes\g^*\to\g^*$.

More generally, consider the case that the $\varphi^{ijk}$ do not vanish. 
The above equations show that $\g^*$ no longer is a Lie algebra. Hence we 
just have a Lie algebra $\g\oplus\g^*$ with an isotropic Lie subalgebra $\g$. 
In this situation, the pair $(\g\oplus\g^*,\g)$ is known as a \emph{Manin pair}. 
Equivalently, the Lie algebra $\g$, supplied with the additional cocommutator 
$\delta$ and $\varphi\in\ext[3]\g$, is said to be a quasi-Lie bialgebra. 
Hence, we find that the solutions of the BV master equation are given 
by quasi-Lie bialgebras. Reversely, any quasi-Lie bialgebra 
(Manin pair) gives rise to a topological open membrane model of the above form. 
The total Lie algebra $\g\oplus\g^*$ is called the Drinfeld double. 

Using this basis we can combine the superfields into a single 
$\Pi(\g\oplus\g^*)$-valued superfield $\BPs=\Bch^i e_i + \Bps_i e^i$. 
Using the canonical inner product $\langle e_i,e^j\rangle=\delta_i^j$, 
we can write the action in the form 
\begin{equation}
  S = \int_\CV \biggl(\frac{1}{2}\langle\BPs,\Bd\BPs\rangle 
   +\frac{1}{3}\langle\BPs,[\BPs,\BPs]\rangle\biggr).
\end{equation}
Noting that the physical fields of ghost number $0$ are the vector components, 
this can be identified with the Chern-Simons theory for the total Lie-algebra 
$\g\oplus\g^*$. This total Lie-algebra is also known as the Drinfeld double. 
Hence, our membrane theory based on the quasi-Lie bialgebra $(\g,\g^*)$ reduces 
to Chern-Simons for the Drinfeld double.

\subsection{Scrooching as a Canonical Transformation}

We now consider the canonical transformation 
\begin{equation}
  \chi^i\to\chi^i+\frac{\del\alpha}{\del\psi_i},
\end{equation}
where the generating function is the degree 2 function 
$\alpha=\frac{1}{2}a^{ij}\psi_i\psi_j$. 
If we started with a quasi-Lie bialgebra, that is $\gamma$ satisfies the 
master equation, we still have a solution to the master equation and 
hence a quasi-Lie bialgebra after this transformation. Note that 
also the boundary term is not affected as $P_\CL\beta=0$. So in fact the 
solution is equivalent. The effect on the structure constants 
$c^i_{jk}$, $f_i^{jk}$ and $\varphi^{ijk}$, are precisely the transformations 
that Drinfeld originally dubbed as twisting of the quasi-Lie bialgebra, 
and is also known as scrooching. 

Similarly, we can consider the canonical transformation 
\begin{equation}
  \psi_i\to\psi_i+\frac{\del\beta}{\del\chi^i},
\end{equation}
with $\beta=\frac{1}{2}b_{ij}\chi^i\chi^j$. Now we have to be careful 
that the boundary term vanishes. If we consider the case where $f=h=0$, 
this implies the condition 
\begin{equation}
  c^{jk}_i\chi^i\frac{\del\beta}{\del\chi^j}\frac{\del\beta}{\del\chi^k} = 0, 
\end{equation}
or in components
\begin{equation}
  f_{i}^{lm}b_{lj}b_{mk}+f_{k}^{lm}b_{li}b_{mj}+f_{j}^{lm}b_{lk}b_{mi}=0.
\end{equation}
This is easily identified with the classical Yang-Baxter equation for 
$b_{ij}$. More generally, we find that $P_\CL\bigl(\e^{-\delta_\beta}\gamma\bigr)=0$ 
can be written 
\begin{equation}
  \frac{1}{2}c_{ij}^k\chi^i\chi^j\frac{\del\beta}{\del\chi^k}
  + \frac{1}{2}f_i^{jk}\chi^i\frac{\del\beta}{\del\chi^j}\frac{\del\beta}{\del\chi^k}
  + \frac{1}{3!}\varphi^{ijk}\frac{\del\beta}{\del\chi^i}\frac{\del\beta}{\del\chi^j}\frac{\del\beta}{\del\chi^k} = 0.
\end{equation}

The boundary observables are functions on the base space $\A=\Pi\g$. 
The space of these functions can be identified with the exterior algebra $\CB=\ext\g^*$. 
The induced $L_\infty$ structure on these boundary observables is given by the 
following differential, bracket, and 3-bracket. 
\begin{eqnarray*}
  d_{\g^*} &=& \frac{1}{2}c_{ij}^k\chi^i\chi^j\frac{\del}{\del\chi^k}, \\
  \{\cdot,\cdot\}_{\g^*} &=& f_i^{jk}\chi^i\frac{\del}{\del\chi^j}\wedge\frac{\del}{\del\chi^k},\\
  \{\cdot,\cdot,\cdot\}_{\g^*} &=& \varphi^{ijk}\frac{\del}{\del\chi^i}\wedge\frac{\del}{\del\chi^j}\wedge\frac{\del}{\del\chi^k}. 
\end{eqnarray*}
Identifying the generators $\beta$ of the above canonical transformations 
with boundary observables, we can write the condition $P_\CL\beta=0$ as 
\begin{equation}
  d_{\g^*}\beta+\frac{1}{2}\{\beta,\beta\}_{\g^*}+\frac{1}{3!}\{\beta,\beta,\beta\}_{\g^*} = 0.
\end{equation}
This is actually a generalization of the quantum Yang-Baxter equation. 

Notice that indeed the structure constants $f_i^{jk}$ determine the 
(Lie) bracket on $\g^*$, and more generally the corresponding Schouten-Nijenhuis bracket 
on the exterior algebra $\ext\g^*$. Also, the structure constants $c_{ij}^k$ of the Lie 
algebra $\g$ induce a cocommutator $d_{\g^*}:\g^*\to\ext[2]\g^*$ on the dual space $\g^*$, 
generalizing to a differential on $\ext\g^*$. This canonical relation between the 
Lie-bracket on $\g$ and a cocommutator on $\g^*$ is well known, and plays an important 
role in the theory of (quasi-)Hopf algebras.

\subsection{Relation to CFT}

It is well known for a long time that Chern-Simons is related to the closed WZW model 
for the same group \cite{witpol}. More recently, it has been shown that also the $G/H$ quotient 
WZW models can be related to Chern-Simons theories. The gauge group of the CS 
in this case $G\times H$. The two gauge fields $A_\pm$ satisfy some nontrivial 
boundary condition, relating the $H$ part of the two gauge fields. For the $G/G$ model, 
this becomes the double CS theory \cite{gawed}
\begin{equation}
  CS(A_+)-CS(A_-) = \int_V\Bigl(A_+dA_+-A_-dA_- + \frac{2}{3}A_+A_+A_+ -\frac{2}{3}A_-A_-A_-\Bigr),
\end{equation}
with the boundary condition $A_+=A_-$. 

We can relate this to our theory. Let $\g$ be a Lie algebra with structure constants 
$c^i_{jk}$, and invariant inner product $\eta_{ij}$. The two sets of gauge fields 
$A^i$ and $B_i$ can then be considered as taking values in the same Lie algebra. 
We identify these with the fields in the double CS above by take the diagonal and 
anti-diagonal gauge fields, 
\begin{equation}
  A_\pm^i = \frac{1}{2}(A^i\pm\eta^{ij} B_j). 
\end{equation}
Indeed, the above double CS action is then equivalent to 
\begin{equation}
  \int_V \Bigl(B_idA^i + \frac{1}{2}c_{jk}^iB_iA^jA^k + \frac{1}{6} c^{ijk}B_iB_jB_k \Bigr),
\end{equation}
and the boundary condition reduces to our boundary conditions $B_i=0$. 
We see that $(\g,[\cdot,\cdot],\delta=0,\varphi)$, where the coassociator 
is related to the structure constants as $\varphi^{ijk}=c^{ijk}$. 
It is well known that the $G/G$ model indeed is related to the quasi-Hopf algebra 
$U_h(\g)$ with nontrivial coassociator related to the structure constants. 

More generally, there are CFT's canonically related to any Manin pair or 
Manin triples \cite{getz1,getz2}. The above $G/G$ model is particular example 
of these models. It seems suggestive that these CFT's could be dual to our 
open membranes relate to Manin pairs. The relation would be similar to the 
one \cite{gawed}. Half of the currents on the boundary 
would be the 1-forms $A^i$. The other currents are of the form $g\inv dg$, 
where $g$ is defined as a Wilson line for $B$ ending on the boundary of the membrane. 
Indeed these would give rise to a current algebra reflecting 
the Lie algebra structure of the double $\g\oplus\g^*$, as in \cite{getz1,getz2}.

\section{Open 2-Branes and Quasi-Lie Bialgebroids}
\label{sec:bialgebroid}

The original open 2-brane model of \cite{js} was that of a pure 
3-form WZ term. This model and some generalizations are discussed 
in this section. They can be related to mathematical objects 
called Courant algebroids, which are studied by mathematicians in 
the context of generalized Dirac quantization. They are also 
known to be related as infinitesimal objects to gerbes.

\subsection{The Canonical Topological Open Membrane}

For our next model, we take for the target superspace  
$\CM=T^*[2](\Pi T^*M)$ for any manifold $M$. 
Notice that this falls in the special class of twisted cotangent bundles 
we have singled out. So we can take for the Lagrangian subspace a section 
$\CL\cong \A=\Pi T^*M$ of this fiber bundle. 

The open membrane theory is defined by four sets of superfields 
$\BX^i$, $\Bch_i$, $\Bps^i$, and $\BF_i$ of ghost degree 0, 1, 1, and 2 respectively. 
$M^i$ and and $\chi_i$ are coordinates on the base $\A=\Pi T^*M$ and $\psi^i$ 
and $F_i$ are coordinates on the fiber. The BV structure is determined by the 
BV bracket 
\begin{equation}
  \int_\CV\biggl(\frac{\del}{\del\BX^i}\wedge\frac{\del}{\del\BF_i}+\frac{\del}{\del\Bch_i}\wedge\frac{\del}{\del\Bps^i}\biggr).
\end{equation}

The BV action functional will be given by 
\begin{equation}
  S = \int_\CV \Bigl( \BF_i \Bd\BX^i + \Bps^i \Bd\Bch_i + \Bga\Bigr).
\end{equation}
the interaction term $\Bga$ satisfies the master equation $[\gamma,\gamma]=0$ 
and $P_\CL\gamma=0$. The boundary term will be Dirichlet for $\Bps^i$ and $\BF_i$. 
To write down this interaction, we introduce two separate gradings, one for $\chi_i$ 
and one for $\psi^i$ and $F_i$ (the latter will have degree 1). 
Note that these will not be preserved separately. According to these gradings, 
we split the interaction term, $\gamma=\gamma^{3,0}+\gamma^{2,1}+\gamma^{1,2}+\gamma^{0,3}$. 
The most general expressions are 
\begin{eqnarray*}
  \gamma^{3,0} &=& \frac{1}{3!}h^{ijk}(X)\chi_i\chi_j\chi_k,\\
  \gamma^{2,1} &=& -b^{ij}(X)\chi_i F_j+\frac{1}{2}f^{jk}_i(X)\psi^i\chi_j\chi_k,\\
  \gamma^{1,2} &=& a^j_i(X)\psi^i F_j+\frac{1}{2}g_{ij}^k(X)\psi^i\psi^j\chi_k,\\
  \gamma^{0,3} &=& \frac{1}{3!}c_{ijk}(X)\psi^i\psi^j\psi^k.
\end{eqnarray*}
First when $a$ is a invertible matrix, we can always use a canonical transformation 
to make it equal to $a_i^j=\delta_i^j$. In the following we will assume this is the case. 
In general, we can transform $a$ to $aU$ for any invertible matrix $U$. Therefore, 
the only relevant information is the rank of $a$. 
The bulk master equation $[\gamma,\gamma]=0$ then implies the following constraints 
\begin{equation}
  f_i^{jk} = -\del_ib^{jk}+c_{ilm}b^{lj}b^{mk},\qquad
  g_{ij}^{k} = -c_{ijl}b^{lk},\qquad
  h^{ijk} = -b^{l[i}\del_lb^{jk]} - c_{lmn}b^{li}b^{mj}b^{nk},
\end{equation}
and furthermore $\del_{[i}c_{jkl]}=0$. The boundary master equation $P_\CL\gamma=0$ 
constrains $h^{ijk}=0$. In other words, the data is given by a closed 3-form 
$c$ and a bivector $b$, satisfying the above constraint. We note that 
for $c=0$ the constraint $h=0$ says that $b^{ij}$ is a Poisson bivector. Hence we 
have a deformed version of the a Poisson bivector, also called a quasi-Poisson structure. 
We will later see that there is a gauge transformation which changes $c$ by an exact form, 
so that actually the data is a 3-form class and a quasi-Poisson structure for a 
representative of this class. 

Combining the above, the total action can be written 
\begin{eqnarray}
S &=& \int_\CV \Bigl( 
\BF_i \Bd\BX^i + \Bps^i \Bd\Bch_i + \BF_i\Bps^i 
 - \Bb^{ij}\BF_i\Bch_j - \frac{1}{2}\del_k\Bb^{ij}\Bps^k\Bch_i\Bch_j 
 + \frac{1}{2}\Bb^{il}\del_l\Bb^{jk}\Bch_i\Bch_j\Bch_k 
\nonumber\\ &&\phantom{\int_CV} 
 + \frac{1}{6}\Bc_{ijk}(\Bps^i-\Bb^{il}\Bch_l)(\Bps^j-\Bb^{jm}\Bch_m)(\Bps^k-\Bb^{kn}\Bch_n)
\Bigr).
\end{eqnarray}

The above action can be derived from the deformation by 
$\e^{-\delta_\beta}\gamma$, where 
\begin{equation}
\gamma = \psi^iF_i+\frac{1}{6}c_{ijk}\psi^i\psi^j\psi^k,\qquad
\beta=\frac{1}{2}b^{ij}\chi_i\chi_j. 
\end{equation}
It is therefore equivalent to the bulk/boundary action 
\begin{eqnarray}
S &=& \int_\CV \Bigl( 
\BF_i \Bd\BX^i + \Bps^i \Bd\Bch_i + \BF_i\Bps^i 
 + \frac{1}{6}\Bc_{ijk}\Bps^i\Bps^j\Bps^k
\Bigr)
+\int_{\del\CV}\frac{1}{2}\Bb^{ij}\Bch_i\Bch_j. 
\end{eqnarray}
We have to be careful however that in the latter case the Lagrangian embedding 
$L\subset\CM$ is not the zero section, but rather is determined by the equations 
\begin{equation}
  \psi^i=[\beta,\psi^i]=b^{ij}\chi_j,\qquad 
  F_i=[\beta,F_i]=\frac{1}{2}\del_kb^{ij}\psi^k\chi_i\chi_j. 
\end{equation}
This affects the projector $P_\CL^\beta$, and therefore the boundary master equation 
$P_\CL^\beta\gamma=0$. 

This membrane action is classically equivalent to the membrane coupling to the $c$-field 
through the WZ term $\int_V c$, and the (closed) Cataneo-Felder on the boundary. 
Therefore, it can be interpreted as a deformation of the Cataneo-Felder model 
by the 3-form.

\subsection{Deformations of the 2-Algebras of Polyvector Fields}

The boundary algebra $B$ of this model has the form $C^\infty(\Pi T^*M)=\Gamma(\ext TM)$. 
In other words, it is the exterior algebra of polyvector fields. These 
are written as functions of $X^i$ and $\chi_i$. The $\chi_i$ can indeed 
be seen as a basis of vector fields. The product 
in this algebra is the wedge product in this exterior algebra. 
The bracket on this algebra for $c=0$ is given by 
\begin{equation}
  \{\cdot,\cdot\} = \frac{\del}{\del X^i}\wedge\frac{\del}{\del\chi_i}.
\end{equation}
This bracket on the algebra of polyvector fields is well known, and is the 
Schouten-Nijenhuis bracket and is the extension to 
polyvector fields of the Lie bracket on vector fields. With this structure, 
the algebra of polyvector fields is well known to be a Gerstenhaber or 2-algebra. 

When we turn on $c$ and $b$, the deformed $L_\infty$ algebra 
structure is given by 
\begin{eqnarray}
Q &=& b^{ij}\chi_j\frac{\del}{\del X^i} 
 + \frac{1}{2}(\del_kb^{ij}+c_{klm}b^{li}b^{mj})\chi_i\chi_j\frac{\del}{\del\chi_k}
 +\CO(c^2), \nonumber \\
\{\cdot,\cdot\} &=& \frac{\del}{\del X^i}\wedge\frac{\del}{\del\chi_i} 
 + \frac{1}{2}c_{ijk}b^{kl}\chi_l\frac{\del}{\del\chi_i}\wedge\frac{\del}{\del\chi_j}
 +\CO(c^2), \\
\{\cdot,\cdot,\cdot\} 
  &=& \frac{1}{6}c_{ijk}\frac{\del}{\del\chi_i}\wedge\frac{\del}{\del\chi_j}\wedge\frac{\del}{\del\chi_k}
 +\CO(c^2). \nonumber
\end{eqnarray}
This forms a $G_\infty$ algebra when we take into account the 
(canonical) product.

\subsection{Canonical Transformations and Gauge Transformations}

In this subsection, we look at space-time  gauge transformations in the topological open 
membrane. In the worldvolume, they correspond to adding BRST exact terms to the action. 
Actually, in the BV formalism there are corrections to this statement, as the BRST 
exact terms only represent the infinitesimal gauge transformations. 

We are interested in the generalization of a gauge symmetry of the form $c\to c+da$, 
where $a$ is a 2-form. Let us start with the situation $b=0$. Then we can add an 
exact term of the form $\BQ\Bal$, where 
\begin{equation}
  \Bal = \frac{1}{2}\Ba_{ij}\Bps^i\Bps^j.
\end{equation}
This term can be written $\frac{1}{6}(d\Ba)_{ijk}\Bps^i\Bps^j\Bps^k+\Bd\Bal$. 
The last term vanishes, due to the boundary condition of $\Bps$. 
It can easily be confirmed that there are no higher order corrections. 
Therefore this term exactly generates the space-time gauge transformation $c\to c+da$. 

When $b\neq0$, there are corrections involving $b$. With the description of 
canonical transformations above they are are not too difficult to write down. 
We noticed above that when $P_\CL\beta\neq 0$, the canonical transformation 
generated by $\beta$ is not a symmetry due to the boundary term. 
Such transformations generate a series of theories, which, as we saw above, 
are equivalent to adding a boundary term. These can 
be understood in terms of the Goldstone modes of the broken symmetry.
The remaining (space-time) symmetries are generated by $\alpha\in \CA_0=\ker P_\CL$. 

In order to find the corrections for nonzero $b$, we have to be a bit more careful 
in the analysis of the canonical transformation generated by $\alpha$ above.
The first step is to find two functions $\beta'\in \CB$ and $\alpha'\in \CA_0$ 
such that 
\begin{equation}
  \e^{\delta_{\alpha}}\e^{\delta_\beta} = \e^{\delta_{\beta'}}\e^{\delta_{\alpha'}}. 
\end{equation}
The solution has the form 
\begin{equation}
  \alpha'=\frac{1}{2}a'_{ij}\psi^i\psi^j - {a''}_i^j\psi^i\chi_j,\qquad
  \beta'= \frac{1}{2}{b'}^{ij}\chi_i\chi_j, 
\end{equation}
with $b'=b(1+ab)\inv$. This can be shown as follows. First, replace 
$\alpha$ by $t\alpha$. Note that $\beta'$ and $\alpha'$ depend on $t$, therefore 
we denote them $\beta'_t$ and $\alpha'_t$ respectively. 
For $t=0$ we clearly have $\beta'_0=\beta$. 
We still take $\beta'_t$ in the above form, with $b'$ now depending on $t$. 
We then find 
\begin{equation}
  \frac{d}{dt}\Bigl(\e^{-\delta_{\beta'_t}}\e^{t\delta_{\alpha}}\e^{\delta_\beta}\Bigr) 
  = \e^{-\delta_{\beta'_t}}\delta_{\alpha}\e^{t\delta_{\alpha}}\e^{\delta_\beta}
    -\delta_{\dot\beta'_t}\e^{-\delta_{\beta'_t}}\e^{t\delta_{\alpha}}\e^{\delta_\beta}
  = \delta_{\gamma_t}\e^{-\delta_{\beta'_t}}\e^{t\delta_{\alpha}}\e^{\delta_\beta},
\end{equation}
where 
\begin{equation}
  \gamma_t = \e^{-\delta_{\beta'_t}}(\alpha)-\dot\beta'_t = \frac12a(\psi-b'_t\chi)^2-\frac12\dot b'_t\chi^2.
\end{equation}
We need $\gamma_t$ to be in $\CA_0$, which means that the $\chi^2$ term vanishes. 
This gives a differential equation for $b'_t$, which is solved by $b'_t=b(1+tab)\inv$. 
Setting $t=1$ gives back our solution above. To solve for $\alpha'$, we have to solve 
\begin{equation}
  \frac{d}{dt}\e^{\delta_{\alpha'_t}} = \delta_{\gamma_t}\e^{\delta_{\alpha'_t}}.
\end{equation}
As $\alpha'_t$ now depends on $t$, this will be a rather complicated 
differential equation. Luckily we will not need the explicit solution; 
the only relevant fact is that the solution for $\alpha'$ is in $\CA_0$, 
which therefore has the form given above. For this it was necessary 
that $\gamma_t$ vanishes on the boundary. 

Using this relation, we have the following relations between pure bulk actions 
\begin{equation}
  \int(\tau+\e^{-\delta_\beta}\gamma) \sim \e^{\delta_{\alpha'}}\int(\tau+\e^{-\delta_\beta}\gamma\Bigr)
  = \int(\tau+\e^{-\delta_{\beta'}}\e^{\delta_\alpha}\gamma).
\end{equation}
Note that the generator $\alpha'$ of the canonical transformation that is used 
vanishes on the boundary, and therefore does not change the boundary conditions: it is 
a true canonical transformation on the bulk. 
The left hand side is equivalent to $S_0+\int\gamma\oplus  P_\CL\beta$, 
while the last expression is equivalent to $S_0+\int\e^{\delta_\alpha}\gamma\oplus P_\CL\beta'$, 
so that we have established the equivalence of the two.

The conclusion of this is that the symmetry generated by $\alpha$ on the total 
algebra $\bar\CA$ is given by 
\begin{equation}
  c\to c+da,\qquad
  b\to b(1+a b)\inv. 
\end{equation}
When both $b$ and $1+ab$ are invertible, we can write the latter as 
$b\inv\to b\inv+a$. As a consistency check, it can be shown that 
the combined transformation is a symmetry of the master equation, 
both $dc=0$ and $b\del b+b^3 c=0$. When $b$ is invertible, 
the invariance of the latter can be seen by writing it is 
$d(b\inv)=c$. We will denote the transformation of $b$ by 
${}^{[a]}b\equiv b(1+ab)\inv$.

\subsection{Boundary Conditions, Duality, and Large $c$}

In the supergravity of decoupled open membranes ending on $M5$-branes 
an important role is played by a 3-vector rather than the 3-form 
\cite{bebe,gentheta,schaar,bersch}. This relation between a 3-form and 
3-vector is similar to the relation between the bivector and 2-form in 
the open string case \cite{codo,scho,seiwit}. 

We have build our model on the base space $\A=\Pi T^*M$, parametrized by the 
coordinates $(X^i,\chi_i)$. For $b=0$, the Lagrangian $\CL\subset\CM$ is precisely 
the zero section. However by turning on a boundary term $\beta$, we can change 
it to any section of the twisted cotangent bundle $\CM=T^*[2]\A$. It was sometimes 
useful to identify the section $\CL$ with the base space $\A$. Indeed, when 
$\CL$ is always transverse to the fiber, the projection to the base gives a 
canonical identification. One can however easily convince oneself that 
there is however no need for the section to be transverse. In fact, 
it does not even have to be a section. Our description of the 
model then is not really appropriate and could better be arranged differently. 
As we will see this gives rise to some interesting dualities. As an extreme case, 
by changing the section $\CL$ we can smoothly go from the undeformed situation 
$\psi^i=0$ to an $\CL$ determined by $\chi_i=0$ (and $F_i=0$ in both situations). 
This can be seen as a change of constant $b$ from $0$ to $\infty$. 
The latter situation is an example where $\CL$ is indeed not given as a section 
of the cotangent bundle. As it turns out, we can however still describe 
this situation as a section of some twisted cotangent bundle. To see this, 
we note the equivalence $T^*[2](\Pi T^*M)=T^*[2](\Pi TM)$. The Lagrangian 
determined by $\chi_i=F_i=0$ is now given as the zero section of the latter 
way of writing. Geometrically, this has however a completely different interpretation. 
Note that the base space $\A=\Pi T^*M$ is replaced by its dual $\A^*=\Pi TM$. 
The boundary algebra $B$, which at first was the algebra of polyvector 
fields, now has changed into the algebra of differential forms. 
In terms of the fields, we roughly have interchanged $\chi_i$ and $\psi^i$. 
That this goes further even than the identification of the observables 
can be seen by looking at the algebra. In the new situation the undeformed 
algebra has zero bracket and 3-bracket, but has a differential which is 
precisely the De Rham differential. 

We can spell out the duality in some more detail when $M$ is an even dimensional 
manifold. The bivector $b$ is not everywhere invertible, but we can always write 
it as the difference of two invertible bivectors, $b=\eps+(b-\eps)$. 
This allows us to write the action in terms of  bulk and a boundary term given by 
\begin{equation}
  \gamma = \psi F+\eps\chi F+\frac{1}{2}\eps\del\eps\chi^3+\frac{1}{3!}c(\psi+\eps\chi)^3,\qquad
  \beta = \frac{1}{2}(b-\eps)\chi^2.
\end{equation}
The boundary condition for $\psi$ is $\psi^i=(b-\eps)^{ij}\chi_j$. 
As $b-\eps$ is invertible we can actually write this as a dual boundary 
condition for $\chi$ rather than $\psi$, $\chi_i=((b-\eps)\inv)_{ij}\psi^j$, 
and write the boundary term as 
\begin{equation}
  \beta = \frac{1}{2}((b-\eps)\inv)_{ij}\psi^i\psi^j.
\end{equation}
Using a canonical transformation generated by this $\beta$, we can write the 
action in terms of a pure bulk term. The boundary conditions have now however 
changed to $\chi_i=0$. This pure bulk term has the general form above, with 
the matrix $a$ not necessarily equal to $\delta$ anymore. In fact we will now 
argue that in general it is not invertible. Straightforwardly working out the 
canonical transformation shows that the matrix $a$ is given by 
$a = 1+\eps\inv (b-\eps)=\eps\inv b$. The assumption that $b$ is not invertible, 
now is seen to be equivalent to the new $a$ being not invertible. 
However, as we also assumed $\eps$ to be invertible, there is a term $\eps\chi F$. 
Using this we can always by a canonical transformation get rid of the $\psi F$ term. 
This same canonical transformation will also simplify the rest of the bulk terms 
involving $c$. After the canonical transformation the bulk and the boundary term 
will have the following form 
\begin{equation}
  \gamma = \eps\chi F + \frac{1}{3!}c'\chi^3,\qquad
  \beta = \frac{1}{2}b'\psi^2,
\end{equation}
where ${c'}^{ijk} = \eps^{il}\eps^{jm}\eps^{kn}c_{lmn}+\eps^{l[i}\del_l\eps^{jk]}$, 
and $b'=(b-\eps)\inv+\eps\inv$. 
Note that using a canonical transformation we can change $\eps$ to almost any fixed 
--- but invertible --- form we want. 

This duality could be useful for studying the large $c$ limit of the theory. 
We can take $\eps$ very small, such that $c'$ is small in all directions. 
Then the bulk interactions are all small, and we can apply perturbation theory. 
Note that the boundary term has to be treated exactly, as it will always be 
large. in this situation.

More generally, we can try to deform the algebra of differential forms. 
It turns out however that there are no nontrivial global deformations. 
The only nontrivial bulk deformation is still the 3-form deformation by 
$\gamma=\frac{1}{6}c_{ijk}\psi^i\psi^j\psi^k$. However, this does not 
satisfy the boundary condition $P_\CL\gamma=0$. This could be remedied 
by adding a boundary term, but only if $c=db$ for some 2-form $b$. 
The boundary term then is simply given by $\beta=\frac{1}{2}b_{ij}\psi^i\psi^j$. 
However, now the total deformation is BRST exact (or more precisely, it is 
generated by a canonical transformation). We might only get something nontrivial 
if $b$ is not defined globally in the target space, but only on patches. 

The duality between $\A$ and $\A^*$ show that the deformed 2-algebras are equivalent. 
In a sense, the algebra of forms deformed by a 3-vector $c^{ijk}$ can be seen as a 
$c_{ijk}\to\infty$ limit of the deformed algebra of polyvector fields. As it is believed 
that a stack of M5-branes in the presence of a large $c$-field reduces to exactly the 
TOM model we studied for $c_{ijk}\to\infty$, this would lead us to study precisely 
this deformation of the algebra of differential forms. Notice that the physical 
boundary observables are precisely given by 2-forms $f(X,\psi)=\frac{1}{2}B_{ij}(X)\psi^i\psi^j$. 

In \cite{js} it was shown how also the topological A- and B-model could be 
found by taking particular boundary conditions for the open membrane model 
when $M$ is a K\"ahler manifold. In these models, one can smoothly 
interpolate between the boundary conditions for the A- and the B-Model. 
This is suggestive of mirror symmetry.

\subsection{Generalized Topological Open Membranes}
\label{sec:general}

We now discuss a further generalization of the above, based on the general 
form of the target superspace. It combines the Lie bialgebra case and the 
structure of the canonical open membrane. 

To construct the target space of the generalized model we start from 
the total space of a Grassmann bundle $\A=\Pi A$, the twist of a 
vector bundle $A\to M$ over a manifold $M$. Following the construction 
discussed before, we take for the target superspace the twisted cotangent 
space $\CM=T^*[2]\A$. The above model is a special case, with $A=TM$. 
For the Lagrangian subspace we can again take a section $\CL\cong \A$ 
of this fiber bundle. 

The open membrane theory is defined by four sets of superfields 
$\BX^i$, $\Bch_a$, $\Bps^a$, and $\BF_i$ of ghost degree 0, 1, 1, and 2 respectively. 
$X^i$ and $\chi_a$ are coordinates on the base $\A=\Pi A$ and $\psi^a$ 
and $F_i$ are coordinates on the fiber. The BV structure is determined by the 
BV bracket 
\begin{equation}
  \int_\CV\biggl(\frac{\del}{\del\BX^i}\wedge\frac{\del}{\del\BF_i}+\frac{\del}{\del\Bch_a}\wedge\frac{\del}{\del\Bps^a}\biggr).
\end{equation}

The BV action functional will be given by 
\begin{equation}
  S = \int_\CV \Bigl( \BF_i \Bd\BX^i + \Bps^a \Bd\Bch_a + \Bga\Bigr).
\end{equation}
the interaction term $\Bga$ satisfies the master equations $[\gamma,\gamma]=0$ 
and $P_\CL\gamma=0$. The boundary term will be Dirichlet for $\Bps^a$ and $\BF_i$. 
To write down this interaction, we split the ghost number into two separate gradings, 
such that $\chi_a$, $\psi^a$ and $F_i$ have degrees $(1,0)$, $(0,1)$ and $(1,1)$. 
Note that these will not be preserved separately. According to these gradings, 
we split the interaction term, $\gamma=\gamma^{3,0}+\gamma^{2,1}+\gamma^{1,2}+\gamma^{0,3}$. 
The most general expressions are 
\begin{eqnarray*}
  \gamma^{3,0} &=& \frac{1}{3!}h^{abc}(X)\chi_a\chi_b\chi_c,\\
  \gamma^{2,1} &=& b^{ai}(X)\chi_a F_i+\frac{1}{2}f^{bc}_a(X)\psi^a\chi_b\chi_c,\\
  \gamma^{1,2} &=& a^i_a(X)\psi^a F_i+\frac{1}{2}g_{ab}^c(X)\psi^a\psi^b\chi_c,\\
  \gamma^{0,3} &=& \frac{1}{3!}c_{abc}(X)\psi^a\psi^b\psi^c.
\end{eqnarray*}
Without a boundary term, the boundary master equation will set $\gamma^{3,0}=0$. 
The bulk master equation will be a combination of the two cases we studied before. 
In components, the equations are 
\begin{eqnarray*}
  b^{i[a}\del_if_d^{bc]}+f_e^{[ab}f_d^{c]e} &=& 0,\\
  a_{[a}^i\del_ig_{bc]}^d+b^{id}\del_ic_{abc}+\frac{1}{2}g_{[ab}^eg_{c]e}^d+f_{[a}^{de}c_{bc]e} &=& 0,\\
  a_{[c}^i\del_if_{d]}^{ab}+b^{i[a}\del_ig_{cd}^{b]}+f^{e[a}_{[c}g^{b]}_{d]e} &=& 0,\\
  a_{[a}^i\del_ic_{bcd]}+g_{[ib}^ec_{cd]e} &=& 0,\\
  b^{j[a}\del_jb^{ib]} + b^{ic}f_{c}^{ab} &=& 0,\\
  b^{ja}\del_ja^i_b + a^j_b\del_jb^{ia} + a_{c}^if_{b}^{ca} +b^{ic}g^a_{cb} &=& 0,\\
  a^j_{[a}\del_ja^i_{b]} + a_{c}^ig_{ab}^c  + b^{ic}c_{cab} &=& 0.
\end{eqnarray*}
It can be interpreted as a local version of a quasi-Lie bialgebra. This can be called 
a quasi-Lie bialgebroid. The structure it gives is also known in the mathematical 
literature as a Courant algebroid, reviewed in the next section. 

To see this more precisely, denote by $[\cdot,\cdot]_0$ the BV bracket in the 
fiber direction only. Note that this is exactly the same bracket as for the 
quasi Lie-algebra $\g\oplus\g^*$ case. Correspondingly we write the 
function $\gamma$ as $\gamma_0+\gamma_1$, where $\gamma_0$ does not involve $F$ 
and $\gamma_1$ is linear in $F$. We write 
\begin{equation}
  \delta=[\gamma_1,\cdot]-[\gamma_1,\cdot]_0=(a_a^i\psi^a+b^{ia}\chi_a)\frac{\del}{\del X^i}. 
\end{equation}
We can then write the full master equation in the form 
\begin{equation}
  \delta\gamma_0+[\gamma_0,\gamma_0]_0 = 0,\qquad \delta^2+[\gamma_0,\delta]_0=0.
\end{equation}
Note that the first equation is very much like a field strength, while the second 
equation is a first order differential operator (as the second order part in $\delta^2$ 
is trivially zero). The first equation is a local generalization of the 
Jacobi-like identities for the structure constants $f,g,c$ of the ``quasi-Lie bialgebra'' 
in the fiber. Note also that $\delta$ is a first order differential operator 
with values in the ``quasi-Lie bialgebra''. These equation suggest that we should 
understand $\delta+[\gamma_0,\cdot]_0$ as a covariant connection in the ``quasi-Lie bialgebra''.

Special solutions arise when we take the trilinear terms to be constant. Then the 
fiber has the structure of a genuine quasi-Lie bialgebra. 
If in addition $a$ and $b$ are constants, the second equation says that $\delta$ takes 
it values in the center of this quasi-Lie bialgebra. When the structure constants are 
not constant, this will be modified as above.

\section{Courant Algebroids and Gerbes}
\label{sec:courant}

In this section we shortly discuss the mathematical structure of 
Courant algebroids and its relation to the topological open membrane.

\subsection{Courant Algebroids}

Algebroids are objects that interpolate between Lie algebras and tangent bundles. 
A \emph{Lie algebroid} over a manifold $M$ is a bundle bundle $A$ over $M$ provided with a 
Lie-bracket $[\cdot,\cdot]_A$ on the space of sections and a map $a:A\to TM$ to 
the tangent space called the \emph{anchor}. This map should intertwine the Lie-bracket 
on $A$ and on vector fields, that is $a([X,Y]_A)=[a(X),a(Y)]_{TM}$. When $M$ is a point, 
a Lie algebroid is the same thing as a Lie algebra. Another special Lie algebroid is the 
tangent space itself, where we take for $a$ the identity map and for $[\cdot,\cdot]_A$
the canonical Lie bracket on vector fields. 

Now let us turn to the case of our main interest. 
A \emph{Courant algebroid} is a vector bundle $E\to M$ with a pseudo-Euclidean inner product 
$\langle\cdot,\cdot\rangle$ together with a bilinear operation $\circ$ on $\Gamma(E)$ 
and a bundle map $\rho:E\to TM$, called the anchor, satisfying the following properties 
\begin{enumerate}
\item $e\circ (e_1\circ e_2) = (e\circ e_1)\circ e_2 + e_1\circ (e\circ e_1)$, 
\item $\rho(e_1\circ e_2) = [\rho(e_1),\rho(e_2)]_{TM}$,
\item $e_1\circ (f e_2) = f e_1\circ e_2+(\rho(e_1)\cdot f)e_2$,
\item $\langle e,e_1\circ e_2+e_2\circ e_1\rangle = \rho(e)\langle e_1,e_2\rangle$,
\item $\rho(e)\langle e_1,e_2\rangle=\langle e\circ e_1,e_2\rangle+\langle e_1,e\circ e_2\rangle$
\end{enumerate}
Here $e,e_1,e_2\in\Gamma(E)$ are sections of $E$ and $f$ is a function on $M$. 
Sections of $E$ naturally act on sections of $E$ through left 
multiplication $e\circ$, and on functions through $\rho$. Properties 1, 3 and 5 
say that this action is a derivation of all the products and the inner product. 
Note that the first property becomes the Jacobi identity when the product $\circ$ 
is antisymmetric. This generalization of a Lie algebra is called a Loday algebra. 
The second and third property are similar to that for the anchor map of a Lie 
algebroid. Property 3 shows that the product $\circ$ acts as a first order differential 
operator. The fourth property shows that the symmetric part of the product 
is in some sense ``infinitesimal''. If we introduce the operator 
$\CD:C^\infty(M)\to\Gamma(E)$ defined by $\langle e,\CD f\rangle=\rho(e)f$, 
we can write the symmetric part as $e_1\circ e_2+e_2\circ e_1=\CD\langle e_1,e_2\rangle$. 
Equivalently, the properties above can be expressed in terms of the 
skew-symmetrization $\{e_1,e_2\}=\frac{1}{2}(e_1\circ e_2-e_2\circ e_1)$, 
as was done the original formulation of Courant algebroids \cite{wein,royt}. 
The Jacobi identity for this skew-symmetric bracket has an anomaly, as the first 
property becomes $\{e_1,\{e_2,e_3\}\}+cycl. + \CD\{e_1,e_2,e_3\}=0$, 
where $\{e_1,e_2,e_3\}=-\frac{1}{6}\langle\{e_1,e_2\},e_3\rangle+cycl$. 
Together with the other identities, these structures give rise to a structure 
of homotopy Lie algebra on the total space of sections and functions \cite{royt}. 

When $M$ is a point, the definitions above reduce to that of (the double of) 
a quasi-Lie bialgebra. Indeed, as $\rho=0$ the product is skew-symmetric, and reduces 
to a genuine Lie bracket on the double. 

A special and canonical example is the so called exact Courant algebroid. 
This is an extension of the tangent bundle, which can be described locally 
as $E\cong TM\oplus T^*M$. The product, or Courant bracket, is defined 
as an extension of the Lie-bracket of vector fields by the formula 
\begin{equation}\label{courexact}
  (v,\xi)\circ(w,\eta) = 
 \Bigl([v,w],\CL_v\eta-\iota_wd\xi+\iota_v\iota_wc\Bigr),
\end{equation}
where $c$ is a closed 3-form and $\iota$ denotes contraction. With the canonical 
inner product $\langle(v,\xi),(w,\eta)\rangle=\iota_v\eta+\iota_w\xi$ 
and anchor $\rho(v,\xi)=v$ one can check that this is indeed a Courant algebroid. 
An exact Courant algebroid can be defined globally as an extension 
\begin{equation}
  0\to T^*M\to E \to TM\to 0. 
\end{equation}
A particular choice of splitting $TM\oplus T^*M$ is called a connection. 
The difference of two such connections can be identified locally with a 2-form. 
The corresponding curvature of the connection, which can be identified 
locally with the exterior derivative of the 2-form connection, is a 
globally defined closed 3-form. In fact this 3-form can be identified 
with the 3-form $c$ appearing in the product above. 
In this way exact Courant algebroids can be classified by a 3-form class. 

A more general class of Courant algebroids can be defined as extensions 
of Lie algebroids of the form 
\begin{equation}
  0 \to A^* \to E \to A \to 0,
\end{equation}
where $A$ and $A^*$ are dual Lie algebroids, possibly with 
trivial anchor and bracket. We can locally split this 
bundle as $E=A\oplus A^*$. The Courant algebroid structure 
on $A\oplus A^*$ is a direct generalization of the exact Courant 
algebroid. Note that we can still define a 
contraction $\iota:A\times A^*\to \R_M$. Also, we can 
define a generalization $d:C^\infty(M)\to \Gamma(A^*)$ of the 
De Rham differential, using the anchor map $a:A\to TM$. 
It is given by the composition of the ordinary De Rham differential 
and the adjoint $a^*:T^*M\to A^*$ of $a$. 
The Lie derivative on sections of $A^*$ can then be defined using the 
standard formula $\CL_v=\iota_v d+d\iota_v$ for $v\in\Gamma(A)$. 
This allows us to write down a generalization of the Courant 
bracket \eqref{courexact} for $E=A\oplus A^*$, with $c\in\Gamma(\ext[3]A^*)$ 
satisfying $dc=0$.

We can generally relate the closed topological membrane with target superspace 
$\CM=T^*[2](\Pi A)$ with the Courant algebroid by identifying $E=A\oplus A^*$. 
Sections of $E$ can then be identified as degree one elements of the 
closed membrane algebra $\CA$, $\Gamma(\Pi E)=\CA^1$, 
and functions on $M$ with degree zero elements. 
More explicitly, a section  $e=(v,\xi)\in\Gamma(E)$ is identified with the 
element $v^a(X)\chi_a+\xi_a(X)\psi^a\in\CA$. 
On this subset of $\CA$ the Courant algebroid structure is defined as 
\begin{equation}
  \langle e_1,e_2\rangle = [e_1,e_2],\qquad
  e_1\circ e_2  = [[\gamma,e_1],e_2],\qquad
  \rho(e)f = [[\gamma,e],f],
\end{equation}
where $\gamma$ is the bulk deformation of the BV action. 
Note that on degree one elements the bracket 
$[\cdot,\cdot]$ is symmetric. Also note that we can identify $\CD f=[\gamma,f]=Qf$. 
More generally, we can identify the $L_\infty$ brackets in the skew-symmetric 
formulation with the derived higher brackets as in \eqref{derlinfty}, 
but without the $P_\CL$ and $i_\CL$. 
One easily verifies that these satisfy the above 
conditions for a Courant algebroid, as a result of the master equation. 
We observe that the topological membranes based on $\CM=T^*[2](\Pi T^*M)$ 
give rise to an exact Courant algebroid, while the more general 
topological membranes with target superspace $\CM=T^*[2](\Pi A)$ 
correspond to the more general Courant algebroid $E=A\oplus A^*$. 

It was observed in \cite{royt} that the total space $A\oplus A^*$ of the Courant 
algebroid is not a symplectic manifold. However it can be naturally embedded into the  
symplectic manifold $T^*A\cong T^*A^*\cong A\oplus A^*\oplus T^*M$. 
By twisting, we find precisely 
the target supermanifold $\CM$ of the generalized topological open membrane. 
This symplectic supermanifold naturally appears in the mathematical construction 
\cite{royt}. In fact, Courant algebroids are in 1-to-1 correspondence to 
supermanifolds of this form supplied with the BV structure and the 
BRST operator \cite{royt2}.

\subsection{Dirac Structures}

To extent this to the open membrane, we need to discuss Dirac structures. 
A Dirac bundle is a maximally isotropic subbundle $L\subset E$ 
(with respect to $\langle \cdot,\cdot\rangle$) that is closed under the action of 
the Courant bracket. We will concentrate on the exact case; 
the generalization is straightforward. A canonical choice for $L$ is 
the subbundle $T^*M$. Deformations of this Dirac structures can 
be found in terms of a bivector $b\in\Gamma(\ext[2]TM)$. Such a 
bivector can be identified with a map $\tilde b:T^*M\to TM$. For $c=0$, 
the graph of this map, spanned by elements $(\tilde b(\xi),\xi)\in TM\oplus T^*M$, 
defines a Dirac structure if $[b,b]=0$, where the bracket is the 
Schouten-Nijenhuis bracket, i.e.\ $b$ is a Poisson structure. 
For nonzero $c$, the condition is replaced by 
$[b,b]=\tilde b^3c$, where the right-hand side is defined as triple contraction. 
In local coordinates we have $3b^{l[i}\del_lb^{jk]}=b^{il}b^{jm}b^{kn}c_{lmn}$. 
This is precisely the condition on $b$ we found for the master equation. 

Another canonical Dirac bundle is given by the graph of a 2-form $b'$, 
spanned by $(v,\iota_vb')\in TM\oplus T^*M$. In the undeformed case, 
this defines a Dirac structure if $db'=0$. This corresponds to the dual 
boundary conditions $\chi=0$, but with the interaction $\gamma=\psi^iF_i$. 
When we deform by $\frac16c_{ijk}\psi^i\psi^j\psi^k$, we find that the 
condition for a Dirac bundle is changed to $db'=c$. This is different 
from the above situation, for three reasons. The first one is that 
there is not always a global solution; only if $c$ is exact. Next 
the solution is fixed up to trivial terms by $c$, and $b'=0$ is not a solution. 
Lastly, the situation is actually gauge equivalent to the trivial situation 
$b'=c=0$. This was not appreciated in the mathematical context. 
But in our case the fact that this situation is more constraint and 
actually trivial follows from the master equation. The difference 
is the extra condition coming from the boundary term in the master equation. 
As we saw earlier, a more interesting case is to trade in the 
interaction $\gamma=\psi^iF_i$ by $\gamma=\eps^{ij}\chi_iF_j$. This 
does have a nontrivial deformation involving a 3-vector, namely by 
$\frac16c'{}^{ijk}\chi_i\chi_j\chi_k$. 

It is clear that the choice of a Dirac bundle $L$ corresponds to the choice 
of a boundary condition for the open membrane. The precise relation is 
that for a Lagrangian $\CL$ describing the boundary condition the corresponding 
Dirac structure is perpendicular with respect to the inner product 
$\langle\cdot,\cdot\rangle$, that is $\Pi L=\CL^\perp$. Therefore 
the sections of $L$ are identified with the degree one elements in 
the kernel of $P_\CL$, i.e.\ $\Gamma(\Pi L)=\CA_0^1$. 
One can indeed see that the boundary master equation $P_\CL\gamma=0$ 
is equivalent to the closure of the Courant bracket on $L$. 
We already saw how the boundary bracket, and more generally the $L_\infty$ 
algebra of the boundary, was related to the derived bracket. According to 
the identification above, the latter is precisely the Courant bracket. 
The boundary $L_\infty$ algebra is the projection of the $L_\infty$ 
algebra related to the skew-symmetrized Courant bracket $\{\cdot,\cdot\}$ 
mentioned above. Reversely, the Courant algebra structure may be seen 
as the general form of the boundary algebra irrespective of the 
boundary condition. 

As a short aside, let us remark that originally Courant algebroids and 
Dirac structures were discovered in the context of general constraint quantization 
of the manifold $M$ \cite{cour}. Generically, the Dirac bundle $L$ can be 
split into three parts: $L\cap TM$, $L\cap T^*M$, and the rest. 
The first factor consists of pure vectors, and corresponds to gauge transformations. 
The second part corresponds to Casimirs generating constraints, while 
all the rest are ordinary dynamical degrees of freedom. For example, 
in the case of the graph of a Poisson bivector $b$, the Casimirs correspond to 
the kernel of the Poisson bivector. These are indeed the central elements 
for the Poisson bracket. Similarly, for the graph of a closed 2-form $b'$, 
the vectors in the kernel of $b'$ generate gauge symmetries for the system. 
$b'$ then becomes a symplectic structure on the quotient manifold of the 
corresponding foliation. More general Dirac bundles can combine both effects. 
This story applies to zero deformation. When we turn on $c$ we deform the 
quantization procedure. However, we see that we still have a well defined 
notion of (integrable) Dirac bundles. This will allow us to perform a quantization of $M$. 
What is the precise meaning of the twisting by $c$ is not completely clear yet, 
however. This involves the quantization of the Courant algebroid.

\subsection{Gerbes and Local Star Products}

Courant algebroids have are closely related to abelian gerbes. This is already 
suggested by the fact that both the exact Courant algebroid and the abelian 
gerbe is classified by a 3-form class. One can be more specific than this. 
An important role in the connection is played by the gauge transformations 
we found above. The data for the model is encoded by the closed 3-form $c$ 
and the boundary bivector $b$, modulo gauge transformations. 
A connection on abelian gerbes is a 2-form $a$. The idea is to identify 
the 3-form $c$ with the curvature of this connection. Thus locally we want to 
write $c=da$. If we could write $c=da$ globally on $M$, we can use the gauge 
transformations to gauge away $c$. The boundary data $B$ will be replaced by 
${}^{[-a]}b=b(1-ab)\inv$. As $c=0$ after the gauge transformation this is a 
genuine Poisson structure. In this case we actually know what quantization 
does: it gives a global star product through deformation quantization. 

For the case that $c$ is globally not exact, let us choose a good covering 
$\{U_\alpha\}$ for $M$. As $c$ is closed, we can on each patch $U_\alpha$ 
choose a 2-form $a_\alpha$ satisfying $c=da_\alpha$. We can then locally 
gauge away $c$ by a gauge transformation generated by $a_\alpha$. 
This gives on each patch a Poisson bivector 
$b_\alpha = {}^{[-a_\alpha]}b=b(1-a_\alpha b)\inv$. On overlaps 
$U_{\alpha\beta}=U_\alpha\cap U_\beta$ the 2-forms $a_\alpha$ differ 
by exact 2-forms, $a_\alpha-a_\beta=d\lambda_{\alpha\beta}$ on $U_{\alpha\beta}$. 
Here we assumed that the patches are chosen such that intersections 
are always contractable. The transition 1-forms $\{\lambda_{\alpha\beta}\}$ 
automatically satisfy the cocycle condition 
$d\lambda_{\alpha\beta}+d\lambda_{\beta\gamma}+d\lambda_{\gamma\alpha}=0$
on triple intersections. One easily sees that the local Poisson bivectors 
$\{b_\alpha\}$ are related on intersections by the gauge transformation  
$b_\beta={}^{[d\lambda_{\alpha\beta}]}b_\alpha$ on $U_{\alpha\beta}$. 

We find that the original global data $(c,b)$ of a closed 3-form and a 
quasi-Poisson bivector can be translated into a set of local Poisson 
bivectors $b_\alpha$ and transition 1-forms $\lambda_{\alpha\beta}$ 
satisfying the cocycle condition on triple interactions. 
The 3-form class of $c$ can be recovered from this local data. 
Given $a_{\alpha_0}$ at one patch, the $a_\beta$ in any other patch 
are determined by the transition 1-forms $\lambda_{\alpha\beta}$. 
The uniqueness of these is guaranteed by the cocycle condition of 
the $\lambda_{\alpha\beta}$. This determines a 3-form $c=da_\alpha$, 
which is globally defined because the $a_\alpha$ differ by exact forms 
on intersections. Note that this determines $c$ only determined modulo 
an exact form, due to the fact that we had to make an initial choice $a_{\alpha_0}$. 
But we indeed find that the local data is classified by the 3-form class of $c$. 
In addition, from the local data we can recover the quasi-Poisson bivector $b$, 
which also should be globally defined. Hence we are able to recover from 
the local data $\{b_\alpha,\lambda_{\alpha\beta}\}$ the complete 
global data $\{c,b\}$ up to global gauge transformation. 

The undeformed Courant algebroid is related to (deformation) quantization. 
The 3-form deformation of the Courant algebroid is therefore expected 
to change the deformation quantization. The way in which this occurs can 
now be described as follows \cite{sevwein}. On each patch $U_\alpha$ 
we have the local algebra of functions $\CA_\alpha=C^\infty(U_\alpha)$. 
Using the above construction we have also a bivector $b_\alpha$. 
We can use $b_\alpha$ and deformation quantization \cite{kon1} 
to construct an associative star product $*_\alpha$ on $\CA_\alpha$ \cite{burwal1}. 
On each intersection $U_{\alpha\beta}$ the Poisson bivectors $b_\alpha$ 
and $b_\beta$ are related by a gauge transformation generated by 
the exact 2-form $d\lambda_{\alpha\beta}$. This implies that the 
two star products $*_\alpha$ and $*_\beta$ restricted to the subalgebras 
$\CA_{\alpha\beta}=C^\infty(U_\alpha\cap U_\beta)$ are equivalent. 
This equivalence is closely related to the Seiberg-Witten map \cite{seiwit}. 
So what we end up with is a set of 
of algebras $(\CA_\alpha,*_\alpha)$, with for any pair $\alpha,\beta$ 
two subalgebras $\CA_{\alpha\beta}\subset\CA_\alpha$ and 
$\CA_{\beta\alpha}\subset\CA_\beta$, which are equivalent deformation 
quantizations, $(\CA_{\alpha\beta},*_\alpha)\simeq (\CA_{\beta\alpha},*_\beta)$. 
These ``noncommutative gerbes'' were recently discussed also in \cite{asbajusch}.

\section{Discussion and Conclusion}
\label{sec:concl}

In this paper we studied a large class of BV actions for topological open membranes. 
We gave a geometric construction of the algebra of bulk and boundary operators. 
The boundary theory has the structure of an homotopy Lie algebra, determined by 
the bulk deformations. This $L_\infty$ algebra is the natural structure of a 
string field theory \cite{zwie}. The main result is that the generic solutions 
to the master equations are given in terms of Courant algebroids, or 
``quasi-Lie bialgebroids''. We should stress the fact that the Courant algebroid 
gives the structure of the boundary $L_\infty$ algebra irrespective of the 
choice of boundary conditions. This means that the boundary algebra 
is given by a projection of the Courant algebroid structure. 

Just as the path integral of topological open strings can give an expansion 
of the (deformation) quantization of function algebras as in \cite{cafe}, 
the path integral for the membrane will give a quantization of the corresponding 
Courant bialgebroids. These structures are much more complicated, mainly due 
to the fact that we are now deforming a string theory on the boundary. 
We only sketched the first order, semi-classical, approximation in this paper. 
Also, we mainly focused on the homotopy Lie algebra structure. While this is 
an important ingredient of the string theory \cite{zwie,homa2}, the boundary 
string has a more intricate structure of homotopy Gerstenhaber algebra 
\cite{kvz,ksv,tomquant}. The quantization problem of Courant algebroids 
should be a generalization of that for the simpler subclass of quasi-Lie bialgebras. 
The quantization of the latter has been solved effectively by \cite{etikaz}, 
and leads as expected to to quasi-Hopf algebras \cite{drin1}. It turns out 
that the full Hopf algebra structure arises naturally only after passing 
to the full homotopy Gerstenhaber structure \cite{tomquant}. 
Therefore we expect that we need to go beyond the $L_\infty$ structure 
to quantize this object. 

A natural and important question that 
arises is whether our topological open membrane model arises as a 
decoupling limit in string theory. For the 3-form model this was argued in 
\cite{bebe}, where it appeared in the context of $M2$-branes ending on 
$M5$-branes. More generally, we expect non-abelian 2-form theories to arise 
in certain little string theories living on 5-branes. 
These may also be related to nonabelian generalizations of Dixmier-Douady gerbes. 
In that sense it is also encouraging that we found a natural extension 
involving non-abelian Lie algebra structures for our model, namely the 
general Courant algebroids. In the case of open strings, the structure of 
gauge theories based on nonabelian 1-form is very similar to the 
one deformed by the $B$-field in noncommutative geometry. 
One might similarly wonder if the structure of nonabelian 2-forms 
will be analogous to that of 2-form theories deformed by the $C$-field. 
In this way, the study of our model based on the exact Courant algebroid 
could learn us about nonabelian 2-forms. 

The deformation of the open membrane by a 3-form also has its effect 
on the corresponding open string theory corresponding to the boundary 
string. We know that the bivector coupling of the string results in 
deformation quantization of the function algebra, as exemplified by 
the Cattaneo-Felder model. The 3-form will deform this quantization 
in a nontrivial way. This is already seen in the fact that the 
Poisson condition for the bivector changes into the quasi-Poisson 
condition. In the last subsection we used the local gauge symmetries 
to write the formal quantization due to such a quasi-Poisson bivector 
as a set of local star-products on patches. One problem with this 
approach is that this only makes sense for formal quantization, i.e., 
viewed as a formal power series in a quantization parameter $\hbar$. 
For \emph{finite} $\hbar$ the star-product algebra can not be localized 
to a patch. 

Another approach which could make more global sense uses the path 
integral quantization of our model. For this we have to model the 
open string, which is nontrivial as the boundary of a membrane itself 
has no boundary. This can be solved by including boundaries with corners, 
and allowing different boundary conditions on various regions of the boundary. 
Let us divide the boundary $\del V$ into two regions. We take the usual boundary conditions 
corresponding to the Lagrangian submanifold $\CL=\Pi T^*M$ for one region, 
but restrict the fields to only $M$ for the other region. The interface 
of the two regions can be viewed as the boundary for region one. 
On the interface live only operators corresponding to functions on $M$, 
as on the boundary of the Poisson-sigma model. Coupling 
the 3-form $c_{ijk}$ to the bulk and a quasi-Poisson bivector $b^{ij}$ 
to region one will induce a nontrivial quantum product on the 
function algebra living on the interface, defined through the path integral. 
For $c=0$ we can completely forget about the bulk and we 
reproduce star-product of  Kontsevich, as in \cite{cafe}. 

Our models naturally have bulk and boundary deformations, which we saw can be 
nontrivially intertwined. We have seen that the bulk algebra $\CA$ can 
be understood as the Hochschild cohomology of the boundary algebra $\CB$. 
In fact, also the boundary deformations can naturally be understood in the 
context of the full deformation complex. In general the deformation complex 
of a $d$-algebra $\CB$ fits in a short exact sequence of the form 
$\CB[d-1]\to\opname{Def}^*(\CB)\to C^*(\CB,\CB)[d]$, 
c.f.\ \cite{kon1}. This induces a long exact sequence in cohomology, 
$\cdots\to H\!H^{i+d-1}(\CB)\to \CB^{i+d-1}\to H^i(\opname{Def}(\CB))\to H\!H^{i+d}(\CB)\to\cdots$.
Let's take our canonical example $\CB=C^\infty(\Pi T^*M)=\Gamma(M,\ext[*]TM)$ 
as a 2-algebra, corresponding to the exact Courant algebroid. 
From the above exact sequence we learn that the space $H^1(\opname{Def}(\CB))$ 
encoding a deformation of this 2-algebra consists of the two 
pieces, $H\!H^3(\CB)=H^3(M)$ and $\CB^2=\Gamma(M,\ext[2]TM)$. Indeed we found 
that for the open membrane with this boundary algebra the deformation 
of the model was precisely encoded in a 3-form class and a bivector. 
We observe that the two components of the deformation complex explicitly reflect 
the bulk and boundary deformations $\gamma$ and $\beta$ of the open membrane.

\section*{Acknowledgements}

It is a pleasure to thank 
Dimitri Roytenberg, Alan Weinstein, Dennis Sullivan, Hong Liu, 
and Ezra Getzler for interesting discussions. 
This work was partly supported by DOE grant DE-FG02-96ER40959.

\end{document}